\documentclass[10pt,aps,prb,reqno,twocolumn,showpacs,superscriptaddress,reprint,nofootinbib]{revtex4-2}
\usepackage[utf8]{inputenc}
\usepackage{amsmath,amssymb,amsfonts,mathtools}
\usepackage{bm,bbm,braket}
\usepackage{dsfont} 
\usepackage{makeidx}
\usepackage{graphicx}
\usepackage{color}
\usepackage[T1]{fontenc}
\usepackage[english]{babel}
\usepackage{changes,enumitem,todonotes}
\usepackage{soul}
\usepackage[mathscr]{euscript}
\usepackage{tikz}
\usepackage[normalem]{ulem}
\usepackage{hyperref,cleveref} 
\hypersetup{colorlinks=true,linkcolor=blue,urlcolor=blue,citecolor=blue}
\usepackage{mathtools} 
\usepackage{soul,xcolor}
\usetikzlibrary{arrows.meta}
\usepackage{float}
\usepackage{placeins}
\usepackage[percent]{overpic}  
\usepackage{ragged2e}
\usepackage{titlesec}
\usepackage{caption}
\usepackage{subcaption}
\usepackage{csquotes}
\usepackage{indentfirst}
\usepackage{setspace}
\usepackage{appendix}
\usepackage{titlesec}
\usepackage{graphicx}

\renewcommand{\eqref}[1]{(\ref{#1})} 

\makeatletter

\renewcommand{\p@subfigure}{\thefigure}

\renewcommand{\fnum@figure}[1]{FIG.~\thefigure.} 
\makeatother

\renewcommand\thesection{\Roman{section}}

\renewcommand\thesubsection{\Alph{subsection}}

\titleformat{\section}  {\normalsize\bfseries\centering\uppercase}
  {\thesection.}{0.4em}{}           

\titlespacing*{\section}{0pt}{14pt}{6pt} 

\titleformat{\subsection}
  {\normalsize\centering\bfseries}
  {\thesubsection.}{0.4em}{}

\titleformat{name=\subsection,numberless}
  {\normalsize\centering\bfseries}
  {}{0.4em}{}
  
\titlespacing*{\subsection}{0pt}{8pt}{4pt} 

\begin{document}

\setstcolor{red}

\title{Thermalization of exact quantum many-body scars in spin-1 \textit{XY} chain under perturbation}

\author{Himadri Halder}
\affiliation{Raman Research Institute, Bangalore 560080, India} \thanks{This work was carried out under the supervision of Prof. Krishnendu Sengupta at Indian Association for the Cultivation of Science.}

\begin{abstract}
Quantum many-body scars are special eigenstates that violate the eigenstate thermalization hypothesis while residing at finite energy density along with thermalizing eigenstates. The spin-1 \textit{XY} model is known to host a family of such exceptional states originating from long-lived quasiparticle excitations that exhibit anomalously low entanglement entropy and long-time periodic revivals, resulting in weak ergodicity breaking. We study the stability of these scarred states against typical U(1) symmetry preserving perturbation in the \textit{XY} chain. While perturbation theory can describe the deformed scar states at small system sizes, finite-size scaling of the perturbation matrix elements indicate that the scars ultimately thermalize in larger chains. Nonetheless, we demonstrate that the long-range order associated with the scars decays under the perturbation, and we estimate the relaxation timescale of oscillatory dynamics in certain local observables to be of order $\lambda^{-2}$, where $\lambda$ is the perturbation strength.
\end{abstract}

\maketitle

\section{Introduction}
\label{sec:intro}
The equilibrium dynamics of isolated quantum systems are governed by the eigenstate thermalization hypothesis (ETH) \cite{eth_deutsch,eth_srednicki,eth_review_rahul,eth_review_christ}. ETH has recently become a central framework for understanding the emergence of macroscopic statistical physics out of quantum mechanical behavior of a system. In brief, ETH states that the reduced density matrix of an eigenstate for a large finite subsystem is equal to the thermal density matrix described by the microcanonical ensemble with effective temperature determined by the energy of the corresponding eigenstate. It further implies that the expectation values of physical observables in highly-excited eigenstates vary smoothly with energy and are insensitive to microscopic details. Consequently, late-time local measurements fail to retrieve information stored in initial steady quantum states. The concepts of ETH are closely tied to quantum chaos, ergodicity and the predictions of random matrix theory \cite{chaos_to_eth,ks_colq,rand_mat}.

However, violations of ETH are frequently observed in integrable systems and many-body localized (MBL) systems. In integrable systems, symmetries and presence of an extensive set of conserved quantities prevent the system from reaching thermal equilibrium \cite{integrable_sys,integrable_rigol,integrable_biroli,integrable_calabrese,integrable_vidmar}. In contrast, the absence of thermalization in many-body localized systems is attributed to disorder and interactions which give rise to emergent quasilocal integrals of motions \cite{eth_review_rahul,mbl_basko,mbl_huse,mbl_altman,mbl_schreiber,mbl_abanin,mbl_pollman,mbl_liom}. The aforementioned phenomena are examples of strong ergodicity breaking where majority of the eigenstates in an energy window do not satisfy ETH. 

Recently, a different class of mechanism, known as quantum many-body scarring has been discovered which leads to weak ergodicity breaking in many-body systems \cite{ryd_scar_first}. Quantum many-body scars (QMBSs) are nonthermal, low-entanglement exact eigenstates that violate \textit{strong} ETH in nonintegrable systems where most of the eigenstates obey ETH. QMBSs are characterized by persistent oscillations and slow dynamics of certain local observables only when the system is initialized in specific states formed from superpositions of the scarred states. As a result, these states can retain coherence for longer time, and expectation values of physical observables deviate from thermal ensemble predictions. Interestingly, QMBSs are not related to any symmetry of the Hamiltonian \cite{ryd_scar_first}, and the number of such states shows algebraic scaling with the system size. Initially discovered in the Rydberg atom simulator \cite{ryd_simul}, scarring phenomena has been explored in a broad class of systems including the Rydberg-blockaded PXP models \cite{ryd_scar_first,ryd_turner,ryd_khemani,ryd_lin,ryd_iadecola,ryd_choi,ryd_ho,ryd_driven}, the AKLT model \cite{aklt_1,aklt_2,aklt_3}, superconducting qubits \cite{sup_1,sup_2}, cold atom setups \cite{cold_atom_1,cold_atom_2,cold_atom_3}, spin-1 models \cite{spin-1_scar,spin-1_XY_fock_space,kitaev_pro}, and various fermionic models \cite{fermion_1,fermion_2,fermion_3,fermion_4,fermion_5,fermion_6}.

The stability of QMBSs has become an active topic of interest, with several studies conducted in the Rydberg PXP Hamiltonian \cite{ryd_pert_chandran,ryd_pert_silva}, deformed $t$-$J$-$U$ model \cite{hubbard_pert}, spin-1 Kitaev chains \cite{kitaev_1,kitaev_2}, and 1D chain of qubits \cite{qmbs_duality}. In this paper, we examine how the scarred states in a spin-1 \textit{XY} chain respond to perturbation, following the framework introduced in \cite{ryd_pert_chandran} for the PXP model. Sec.~\ref{sec:model} provides a brief review of the QMBSs in a spin-1 \textit{XY} chain, originally proposed in \cite{spin-1_scar}, and presents some numerical results for scaling of long-range correlations, and oscillations in local observables. In Sec. \ref{sec:sec_NN_perturb}, we consider a perturbation that breaks the bipartite lattice structure and is off-diagonal in the working $S_z$ basis (in contrast to the diagonal perturbation studied in \cite{ryd_pert_chandran} for diagnosing spin-1 \textit{XY} scars). We demonstrate the resulting hybridization of the QMBSs with thermal eigenstates and discuss the perturbation theory analysis at small system sizes, along with finite-size scaling of the relevant matrix elements and an estimate of the thermalization timescale.

\vspace{-14pt}
\section{Spin-1 \textit{XY} Model with Perturbation}
\label{sec:model}
We consider a one-dimensional spin-1 XY chain, with a next nearest-neighbor exchange as a perturbation for our study. The model has the following Hamiltonian:
\begin{align}
    H = H_0 + \lambda H_p, \label{H}
\end{align}
\noindent with
\begin{align}
    H_0 =&\ J_1\sum_i \left(S^x_iS^x_{i+1}+S^y_iS^y_{i+1}\right) + J_3\sum_i \left(S^x_iS^x_{i+3}+S^y_iS^y_{i+3}\right) \notag \\
    &\quad +\ h\sum_i S^z_i + D \sum_i \left(S^z_i\right)^2    \label{H_0}
\end{align}
\noindent and
\begin{align}
    H_p = \sum_i \left(S^x_iS^x_{i+2}+S^y_iS^y_{i+2}\right)  \label{H_p}
\end{align}
where $S_i^{\alpha}(\alpha=x,y,z)$ are spin-1 operators defined on $i$-th site of a chain of length $L$.~$J_1$ and $J_3$ are the coupling strengths of the nearest-neighbor and third nearest-neighbor exchange interactions respectively, and $h$ denotes the magnetic field. An onsite anisotropy term is also present, which is characterized by $D$, while $\lambda$ represents the strength of the perturbation $H_p$. Although the scarred states in this model exist for both periodic and open boundary conditions and for any $D,h$ \cite{spin-1_scar}; our study presents results for the OBC only.

Both $H$ and $H_0$ have global U(1) symmetry around the $z$ axis. The third nearest-neighbor term in $H_0$ destroys the additional nonlocal SU(2) symmetry shown in Ref.~\cite{nonlocal_su2} for OBCs. All terms in $H_0$ respect the bipartite structure of the chain but $H_p$ does not preserve it.

\subsection*{Exact many-body scars}
\label{subsec:exact_scar}
The model Hamiltonian $H_0$ is known to be non-integrable, with its many-body level spacing statistics characterized by the Wigner–Dyson (WD) distribution \cite{spin-1_scar}. Although the WD level statistics is commonly taken as an indicator of chaotic and ergodic dynamics in quantum systems, it alone does not establish the \textit{strong} ETH which requires all states within a given energy window to satisfy the ETH \cite{strong_ETH}. In certain scenarios, only a weaker form of the ETH may apply, which permits a small set of atypical eigenstates (number of which typically vary algebraically with system size) that deviate from ETH predictions \cite{integrable_biroli,weak_ergod}.   

In Ref.~\cite{spin-1_scar} it has been shown that the following tower of exact \textit{bimagnon} states do not obey the ETH:
\begin{align}
    \ket{\mathcal{S}{}_n} = \mathcal{N} \left(n\right) \left(J^+\right)^n \ket{\Omega} \label{first_NN_scar}
\end{align}
\noindent where $n = 0,..,L$, and $\ket{\Omega} = \bigotimes_r \ket{m_r=-1}$ is the fully polarized vacuum state. $\mathcal{N} \left(n\right)=\sqrt{(L-n)!/n!L!}$ are the normalization factors and,
\begin{equation}
    J^\pm = \frac{1}{2} \sum_{r=1}^{L} e^{ir\pi} (S_r^\pm)^2  \label{bimagnon_operators}
\end{equation}
Intuitively, the $J^+$ operator creates a wavepacket corresponding to a bimagnon excitation with momentum $k=\pi$ on top of the vacuum (no bimagnon) state and repeated actions of this operator create a condensate of bimagnons at the same momentum. A state $\ket{\mathcal{S}{}_n}$ contains such a condensate of $n$ number of bimagnons. These specific types of bimagnon excitations described in Eq.~\eqref{first_NN_scar} have infinite lifetime because they are insusceptible to scattering and decay due to the destructive interference among the disassociation processes of bimagnons into magnons resulted by the Hamiltonian $H_0$ \cite{spin-1_scar_review}. If we write only the XY exchange part of the unperturbed Hamiltonian Eq.~\eqref{H_0} as $(H_0)_{XY}=J_1\sum_i h^1_{i,i+1}+J_3\sum_i h^3_{i,i+3}$; then each such $h^1_{i,i+1}$ and $h^3_{i,i+3}$ term will individually annihilate any $\ket{\mathcal{S}{}_n}$ state. Hence the states $\ket{\mathcal{S}{}_n}$ are frustration-free zero energy eigenstates of both the $XY$ exchange terms in $H_0$ Hamiltonian. They reside in the total magnetization sector $m_n=2n-L$ and are exact eigenstates of $H_0$ with energy $E_n=h(2n-L)+LD$.

\begin{figure}[H]
    \centering
    \begin{subfigure}{\columnwidth}
        \centering
        \begin{overpic}[width=\columnwidth]       {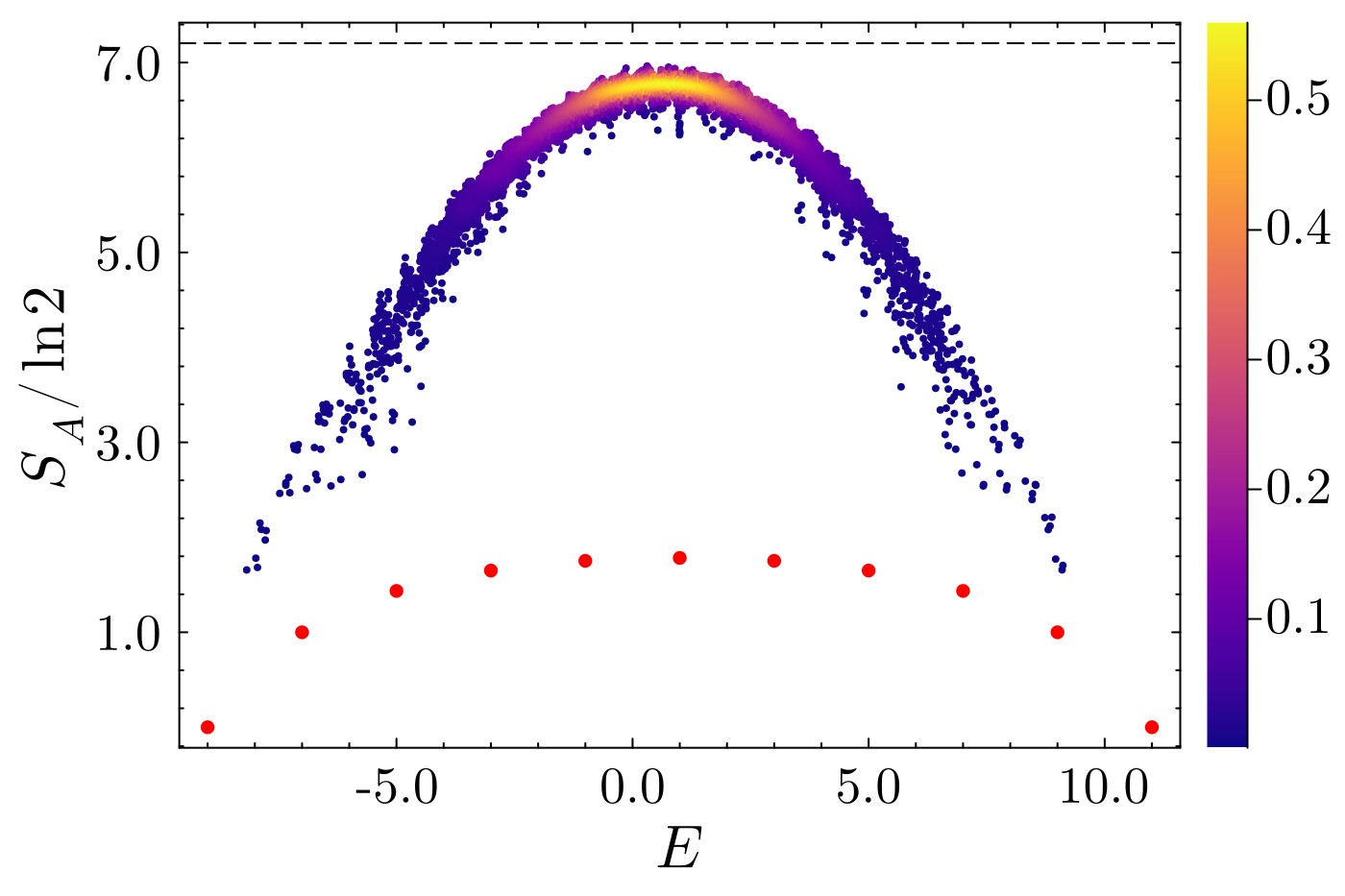}
        \put(13.9,55.8){\large (a)}
        \end{overpic}
        \phantomcaption\label{fig:unp_entang}
    \end{subfigure}%
    \\[-4mm]
    \begin{subfigure}{\columnwidth}
        \centering
        \begin{overpic}[width=\columnwidth]     {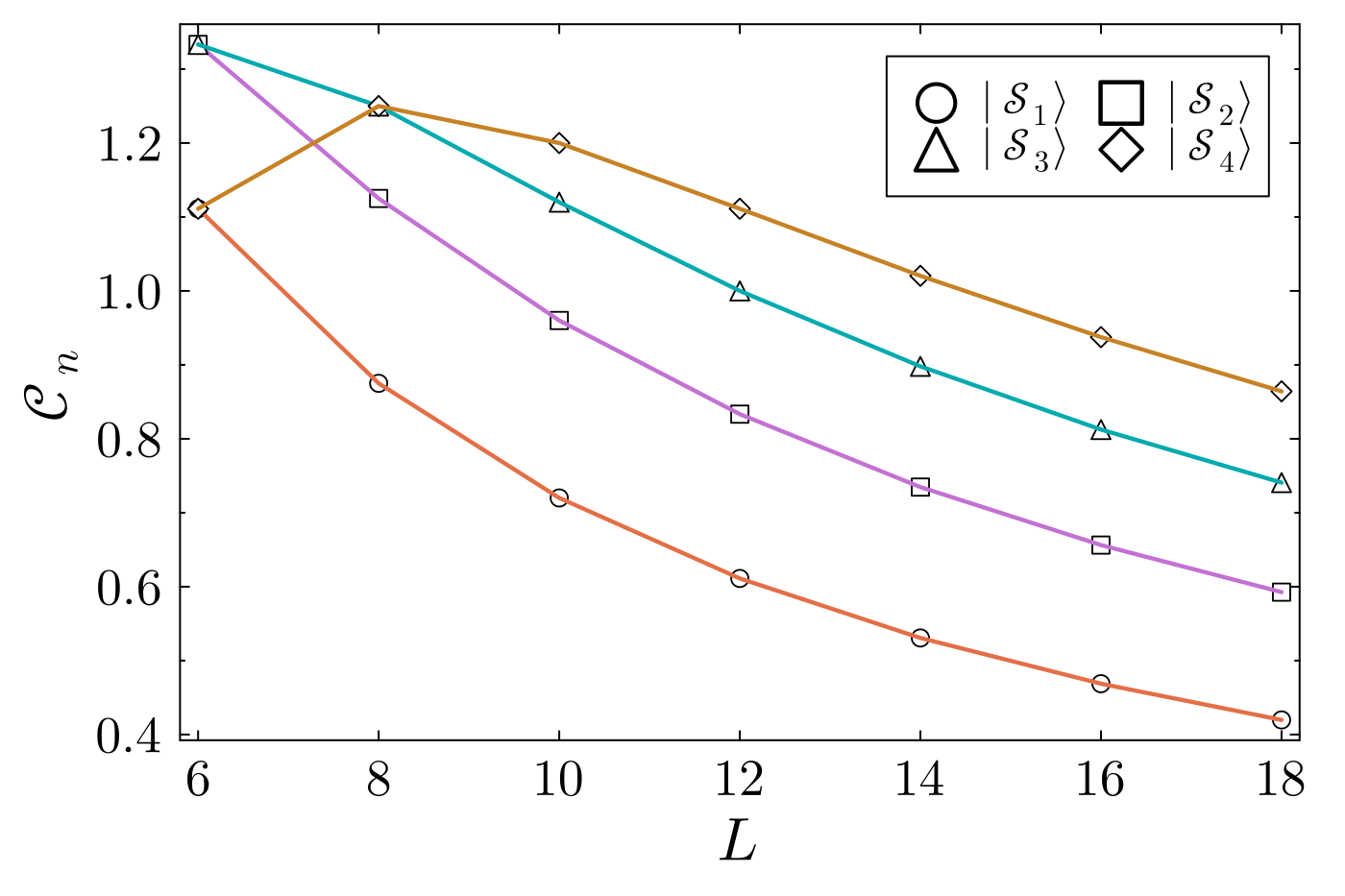}
        \put(13.9,55.8){\large (b)}
        \end{overpic}
        \phantomcaption\label{fig:unp_odlro}
    \end{subfigure}
    \caption{(a) Bipartite entanglement entropy $S_A$ of eigenstates of $H_{0}$ for an $L=10$ spin-1 XY chain with OBC and $(J_1,J_3,D,h)=(1.0,0.1,0.1,1.0)$. Density of states in the zero-magnetization sector is shown by color. Red circles denote $S_A$ for scarred states Eq.~\eqref{first_NN_scar}. The dashed line at the top is the entropy for a random state, $S^{\mathrm{ran}}_A=(L/2)\ln 3-\tfrac{1}{2}$. 
    The scar states are equally spaced in the spectrum and have low entanglement entropy compared to the other eigenstates. 
    (b) Scaling of off-diagonal long-range order ${\mathcal{C}{}_n}$ Eq.~\eqref{odlro} calculated numerically for different scar states, with chain length $L$. The Hamiltonian parameters are the same as in (a). }
    \label{fig:unp_entang_odlro}
\end{figure}

Interestingly, these scarred states in Eq.~\eqref{first_NN_scar} manifest long-range connected correlations which is not present in ETH-following states. The off-diagonal long-range order (ODLRO) \cite{odlro} is demonstrated by the correlation function
\begin{align}
    {\mathcal{C}{}_n} = \bra{\mathcal{S}{}_n}O_{\pi}^{\dagger}O_{\pi}\ket{\mathcal{S}{}_n} = \left[1-\left(\frac{m_n}{L}\right)^2\right]+\mathcal{O}\left(\frac{1}{L}\right) \label{odlro}
\end{align}
where $O_{q}=\left(1/L\right)\sum_{r}e^{irq}(S_r^+)^2$ is a spin-nematic order parameter defined at wave vector $q$. Fig.~\ref{fig:unp_odlro} shows scaling of this long-range correlation with length of the chain. Noticeably, even in the thermodynamic limit $(L\to \infty)$ the correlation survives except the cases when the system has no bimganon or $L$ number of bimagnons. On the contrary, the expectation value of this correlation function calculated for a generic ETH-obeying eigenstate in infinite-temperature ensemble is $(4/3L)$ \cite{spin-1_scar_review} which goes to zero at $L\to\infty$ limit. The existence of such long-range order in condensate of bimagnons at momentum $\pi$ emerges from the fact that these scarred states form a representation of an emergent SU(2) algebra (and also SGA, see Appendix \hyperref[appendix:SGA]{A}) which is generated by the $J^{\pm}$ operators:
\begin{align}
    [J^+,J^-]=2J^z, \hspace{2mm} [J^z,J^{\pm}]=\pm J^{\pm}, \hspace{2mm} J^z=\frac{1}{2}\sum_i S_i^z  \label{su2_generators}
\end{align}
The Hamiltonian Eq.~\eqref{H_0} conserves total magnetization corresponding to the operator $S^z_{\mathrm{tot}}=\sum_i S^z_i$ due to U(1) symmetry generated by spin rotations around the $z$ axis, hence $J^z$ commutes with $H_0$. But $J^{\pm}$ do not commute with $H_0$. Effectively, $J^{\pm}$ are raising and lowering operators for the scarred states while $\ket{\mathcal{S}{}_n}$ are eigenstates of the emergent SU(2) algebra with maximal-spin $L/2$:
\begin{align}
    J^{\pm}\ket{\mathcal{S}{}_n} & = \sqrt{\left(j\mp \frac{m_n}{2}\right)\left(j\pm \frac{m_n}{2}+1\right)}\ket{\mathcal{S}{}_{n\pm 1}} \label{J_low_raise} 
\end{align}
\begin{align}
    J^z\ket{\mathcal{S}{}_n} = \frac{m_n}{2}\ket{\mathcal{S}{}_n}
\end{align}
and
\begin{align}
    \bm{J}\cdot\bm{J}\ket{\mathcal{S}{}_n}=j(j+1)\ket{\mathcal{S}{}_n} \label{su2_eq}
\end{align}
where $j=L/2$, $m_n/2=n-L/2$, and $\bm{J}\cdot\bm{J}=\frac{1}{2}(J^+J^- + J^-J^+)+(J^z)^2$. The result in Eq.~\eqref{odlro} follows directly from Eq.~\eqref{J_low_raise} by noting that $O_{\pi}^{\dagger}O_{\pi}=(4/L^2)J^-J^+$.

Despite having finite energy density and sharing the same symmetry sectors with other states which are governed by ETH, the special tower of states in Eq.~\eqref{first_NN_scar} do not thermalize. This atypical signature can be seen from the entanglement spectrum of the system in Fig.~\ref{fig:unp_entang}. $S_A = -\text{tr}(\rho_A \ln \rho_A)$ is the bipartite Von Neumann entanglement entropy of the eigenstates obtained from exact diagonalization, where $\rho_A$ is the reduced density matrix of the region $A$ of length $L/2$. In Fig.~\ref{fig:unp_entang}, $S_A$ vs energy have been shown for eigenstates in the $m_n=0$ sector. ETH-obeying states follow volume law scaling and entropy of states in the middle of the spectrum should approximately be same as the entropy for a random state, $S^{\mathrm{ran}}_A=(L/2)\ln 3-\tfrac{1}{2}$ \cite{rand_state_entang}. From Fig.~\ref{fig:unp_entang} it appears that this roughly holds for a fairly large number of eigenstates near zero energy region. In contrast, the entropy of $\ket{\mathcal{S}{}_{L/2}}$ which is the most entangled state among all scarred states Eq.~\eqref{first_NN_scar}, varies as $\ln{(\pi L/8)}$ \cite{spin-1_scar} which is an evidence of ETH violation.  

\begin{figure}[H]
    \centering
    \begin{subfigure}{\columnwidth}
        \centering
        \begin{overpic}[width=\columnwidth]       {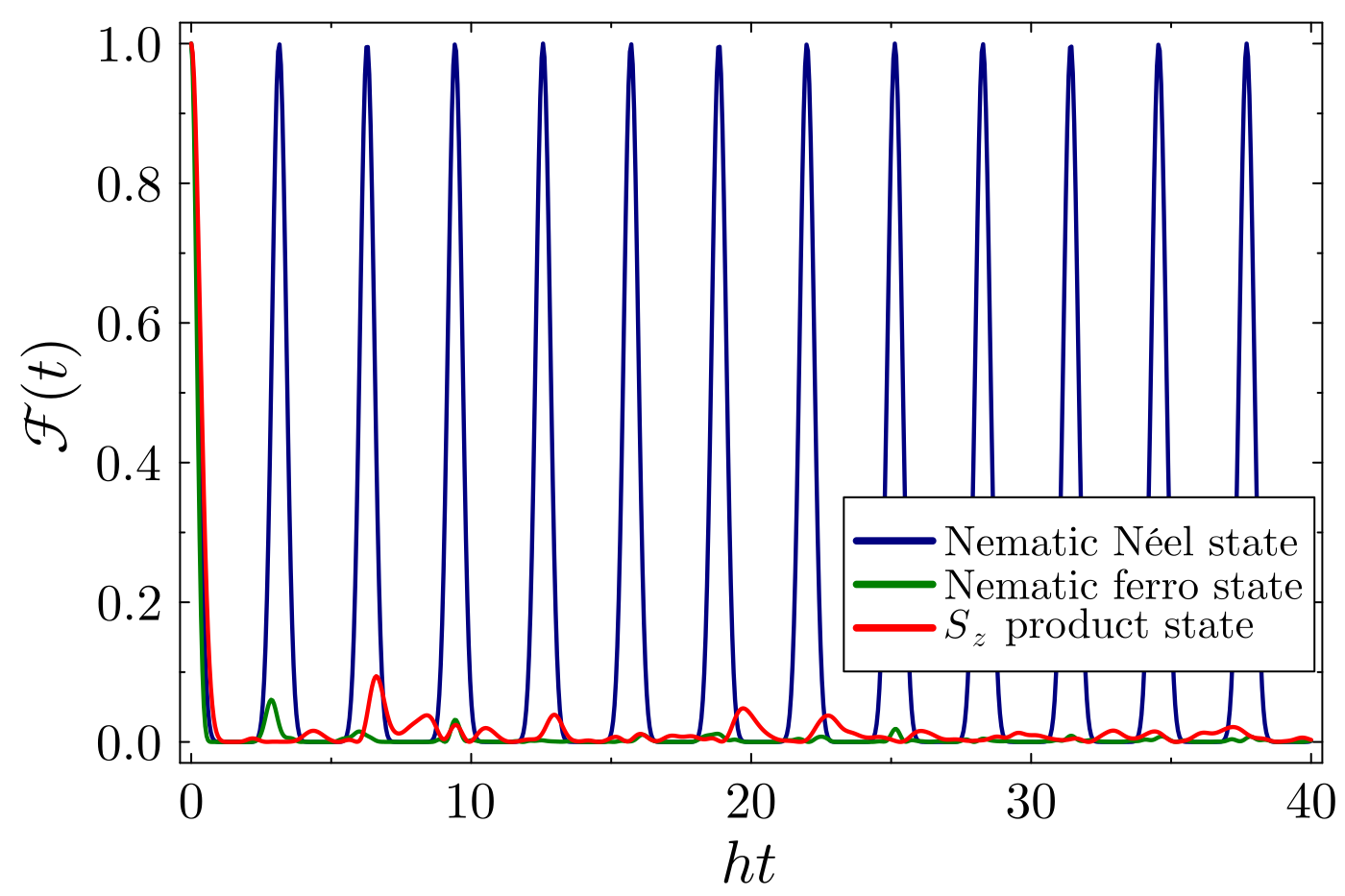}
        \put(14.4,59){\large (a)}
        \end{overpic}
        \phantomcaption\label{fig:unp_fidelity}
    \end{subfigure}%
    \\[-4mm]
    \begin{subfigure}{\columnwidth}
        \centering
        \begin{overpic}[width=\columnwidth]     {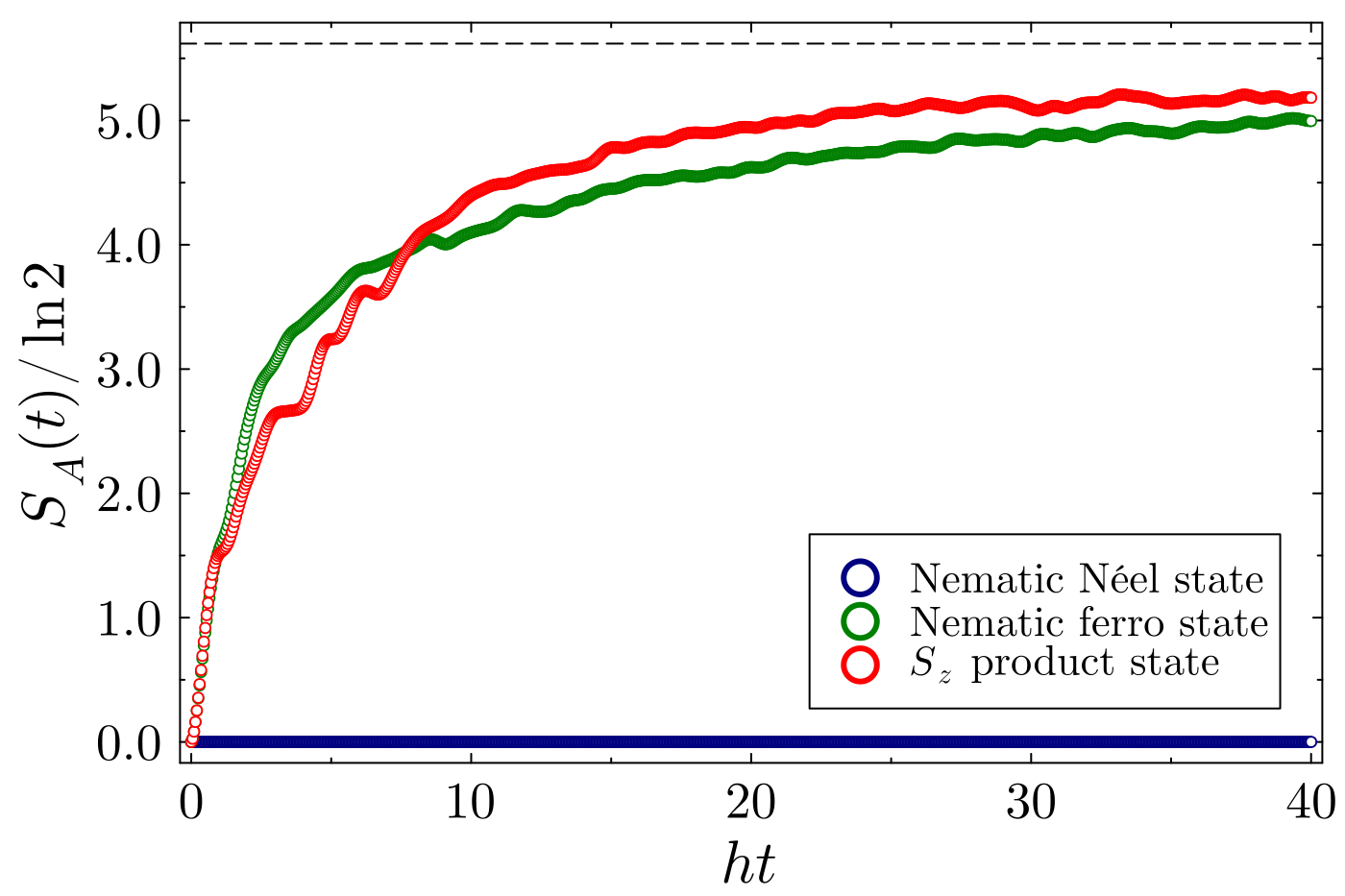}
        \put(14.4,59){\large (b)}
        \end{overpic}
        \phantomcaption\label{fig:unp_entang_evol}
    \end{subfigure}
    \caption{Time evolution of an $L=8$ chain with rest of the parameters identical to those in Fig.~\ref{fig:unp_entang}: (a) Fidelity $\mathcal{F}(t) = |\langle{\psi(0)}|{\psi(t)}\rangle|^2$ for different initial states. The N\'eel state shows perfect periodic revivals in contrast to other non-special states, which display a very fast decay.
    (b) Entanglement entropy of both the nematic ferro state and $S_z$ product state quickly converge to the maximal entropy (shown by the black dashed line) under time evolution with $H_0$, but the N\'eel state shows no change in entanglement dynamics.}   
    \label{fig:unp_fidelity_entang_evol}
\end{figure}

The nonthermalizing nature is also imprinted on the dynamics under the Hamiltonian Eq.~\eqref{H_0} if the system is initialized in certain states. Ref.~\cite{spin-1_scar} has considered a nematic state $\ket{\psi_0}$ in their study to show persistent coherent many-body revivals under Hamiltonian time evolution. The relevance of this particular state lies in the fact that $\ket{\psi_0}$ is the lowest-weight state of $J_x$ in the spin-$L/2$ representation of the SU(2) algebra Eq.~\eqref{su2_generators} because it is the ground state of the Hamiltonian $H_A=\tfrac{1}{2}(J^+ + J^-)\equiv J^x$ with eigenvalue $-L/2$ and is given by,
\begin{align}
    \ket{\psi_0} = \bigotimes_r \left(\frac{\ket{m_r=+1}-e^{ir\pi}\ket{m_r=-1}}{\sqrt{2}}\right) \label{neel_state}
\end{align}
$\ket{\psi_0}$ has no net magnetic moment, i.e., $\langle{S_i^{\alpha}}\rangle=0$ for all sites $i$ where $(\alpha=x,y,z)$ and lacks any dipolar order. This state is referred to as \enquote{nematic N\'eel} state because phase factor of the expectation value of the quadrupolar order parameter $(S_i^+)^2$ in this state alternates at adjacent sites, resulting in a staggered arrangement of spin-nemat-
\noindent ic directors in the \textit{x-y} plane. $\ket{\psi_0}$ remains fully confined to the scarred manifold: $\ket{\psi_0} = \sum_{n=0}^{L}c_n\ket{\mathcal{S}_n}$ with the coefficients given as $c_n^2 = \tfrac{1}{2^L}\binom{L}{n}$. If the system is initialized in this \enquote{N\'eel} state, the subsequent dynamics generated by the Hamiltonian $H_0$ Eq.~\eqref{H_0} is governed by the unitary evolution $\ket{\psi_0(t)}=e^{-iH_0t}\ket{\psi_0}$ and the quantum many-body fidelity is given by
\begin{align}
    \mathcal{F}{}_0(t) = |\langle{\psi_0}|{\psi_0(t)}\rangle|^2 = \cos^{2L}(ht) \label{fidelity}
\end{align} This indicates that the system periodically returns to the initial state with a time period $T=\pi/h$, which suggests violation of ETH, as an ETH-obeying state is not expected to show periodic revivals. In distinction, the nematic ferro state referred to in Fig.~\ref{fig:unp_fidelity} has all the spin-nematic directors aligned in the same direction, rather
\noindent than staggered as in the N\'eel state Eq.~\eqref{neel_state}. It is a product state composed of a local superposition of $\ket{1}$ and $\ket{-1}$ on each site, which has zero bipartite entangle entropy, and is written as, 
\begin{align}
    \ket{\psi_f} = \bigotimes_r \left(\frac{\ket{m_r=+1}+\ket{m_r=-1}}{\sqrt{2}}\right) \label{ferro_state}
\end{align}
As demonstrated in Fig.~\ref{fig:unp_fidelity}, fidelities of other generic states such as the $\ket{\psi_f}$ and an $S_z$ product state exhibit rapid decay. Fig.~\ref{fig:unp_entang_evol} shows entanglement evolution for the same set of states. Interestingly, the special product state $\ket{\psi_0}$ behaves as an exact product state even at late time but other product states rapidly generate entanglement and evolve towards the entropy of a random state (marked by the dashed line). These features of the N\'eel state clearly indicate the absence of thermalization. 

The non-ergodic characteristics are also reflected in the 
\begin{figure}[H]
    \centering
    \includegraphics[width=\columnwidth]{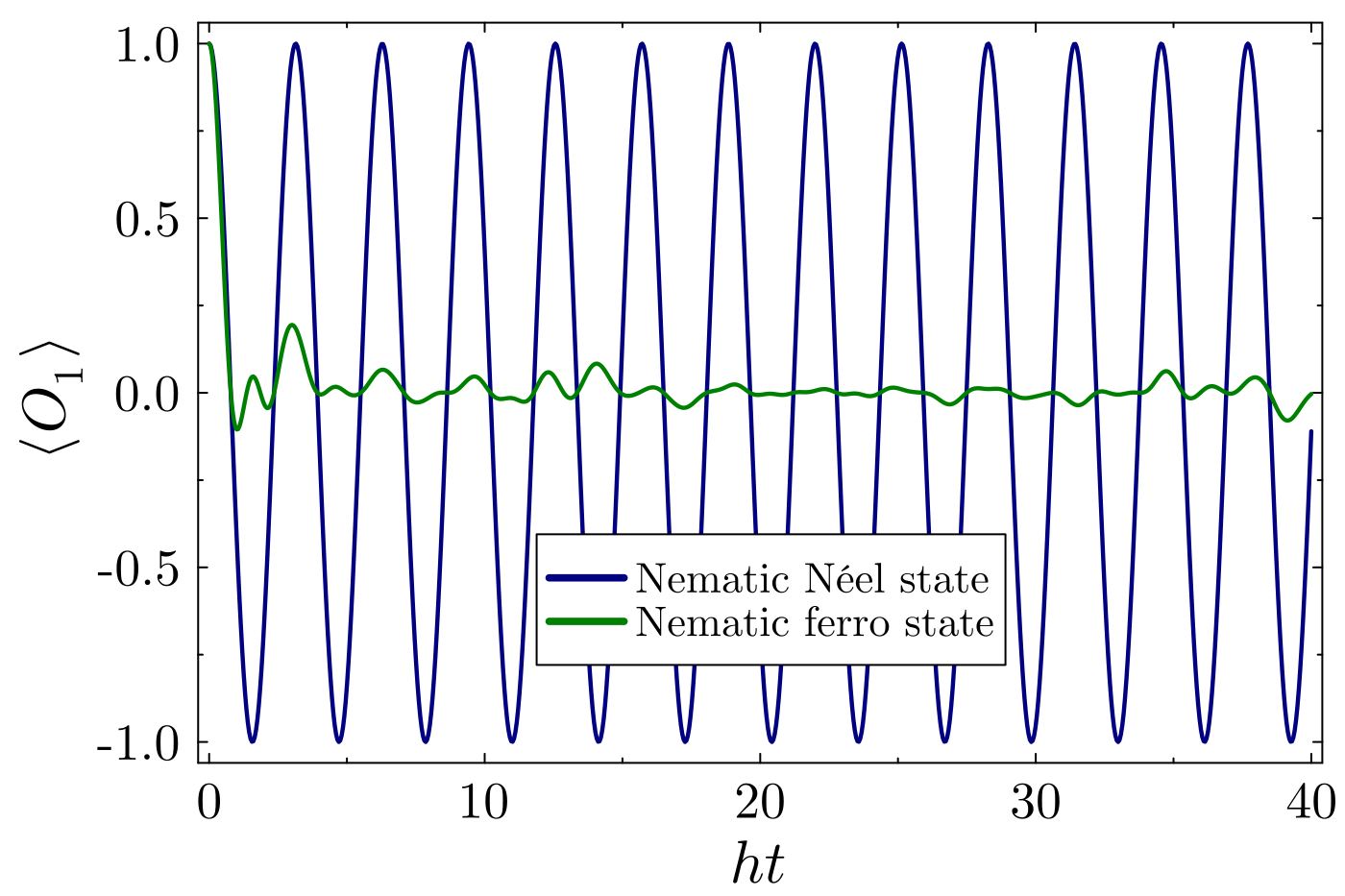}
    \vspace{1mm}
    \caption{Late time dynamics of expectation value of the local observable $O_1$ Eq.~\eqref{nematic_director_param} for both N\'eel and ferro initial states for $L=8$, with the remaining parameters same as in Fig.~\ref{fig:unp_entang}. When initialized in the N\'eel state, perfect oscillations are observed in $\braket{O_1}$ over time $t$, whereas the observable shows no oscillation for the ferro state.}
    \label{fig:unp_director}
\end{figure}
\noindent behavior of the local observable \cite{kitaev_2,spin-1_scar}
\begin{align}
    O_r = \frac{1}{2}\left[(S_r^+)^2+(S_r^-)^2 \right]
    \label{nematic_director_param}
\end{align}
\noindent Phase of $\langle(S_r^+)^2\rangle$ for a nematic state essentially describes the director angle at $r$-th site. Time evolution of the expectation value of this order parameter in the N\'eel state gives,
\begin{align}
    \bra{\psi_0(t)}O_r\ket{\psi_0(t)} = (-1)^{r+1}\cos(2ht)
\end{align} 
\noindent where $r = 1,..,L$. The physical picture associated with the equation above is that the spin-nematic directors in the \textit{x-y} plane rotate around the external field with a frequency of $2h$, i.e., stronger magnetic field leads to faster rotation. As presented in Fig.~\ref{fig:unp_odlro}, following a quench from the N\'eel state with initially staggered directors, all the local directors undergo collective phase-coherent precessional motion. If the directors are instead initially aligned as in the ferro state, the Hamiltonian evolution would lead to rapid decoherence in their dynamics. 

The rotation of the directors can be understood from the fact that the $\ket{\psi_0}$ state can be written as a superposition of scarred eigenstates $\ket{\mathcal{S}{}_n}$, as the N\'eel state stays completely within the scarred manifold. The Hamiltonian dynamics only introduce time-dependent phases to these states while retaining their eigenstate properties. A more concrete origin of the revivals is the emergent SU(2) algebra Eq.~\eqref{su2_eq} which is the underlying source of the long-range order Eq.~\eqref{odlro} in scarred states $\ket{\mathcal{S}{}_n}$, which in turn gives rise to a single emergent staggered director precessing around the applied field. It should be noted that the revivals are perfect in the spin-1 XY chain, i.e., the quantum fidelity returns to $1$ even at late times, owing to the exactness of the embedded SU(2) algebra. In contrast, in the PXP model the revivals are not perfect \cite{ryd_scar_first}, since the SU(2) algebra and the resulting description of a single \enquote{big spin} precessing under the effect of PXP Hamiltonian are merely approximate \cite{ryd_choi,spin-1_scar_review,scar_lie_algebra}.  

\section{Effects of Perturbation on QMBS}
\label{sec:sec_NN_perturb}
The introduction of a next nearest-neighbor interaction $H_p$ Eq.~\eqref{H_p} destroys the bipartite nature of the original spin-1 \textit{XY} Hamiltonian $H_0$ Eq.~\eqref{H_0}. The destructive interference mechanism responsible for the stability of the specially constructed bimagnon states in the unperturbed spin-1 \textit{XY} model fails to hold in presence of the specific types of perturbation discussed in Eq.~\eqref{H_p}. Consequently, the scarred states Eq.~\eqref{first_NN_scar} are no longer exact eigenstates of the full Hamiltonian $H$ Eq.~\eqref{H}. In Sec.~\ref{subsec:exact_scar}, we showed that the periodic revivals in fidelity and the oscillations in local observables arise directly from the eigenstate properties of the scarred states. It is therefore intriguing to investigate how the perturbation effects these atypical features of the scarred states, which distinguish them from the behavior exhibited by generic ETH-obeying states. 

\begin{figure}[H]
    \centering
    \begin{subfigure}{\columnwidth}
        \centering
        \begin{overpic}[width=\columnwidth]       {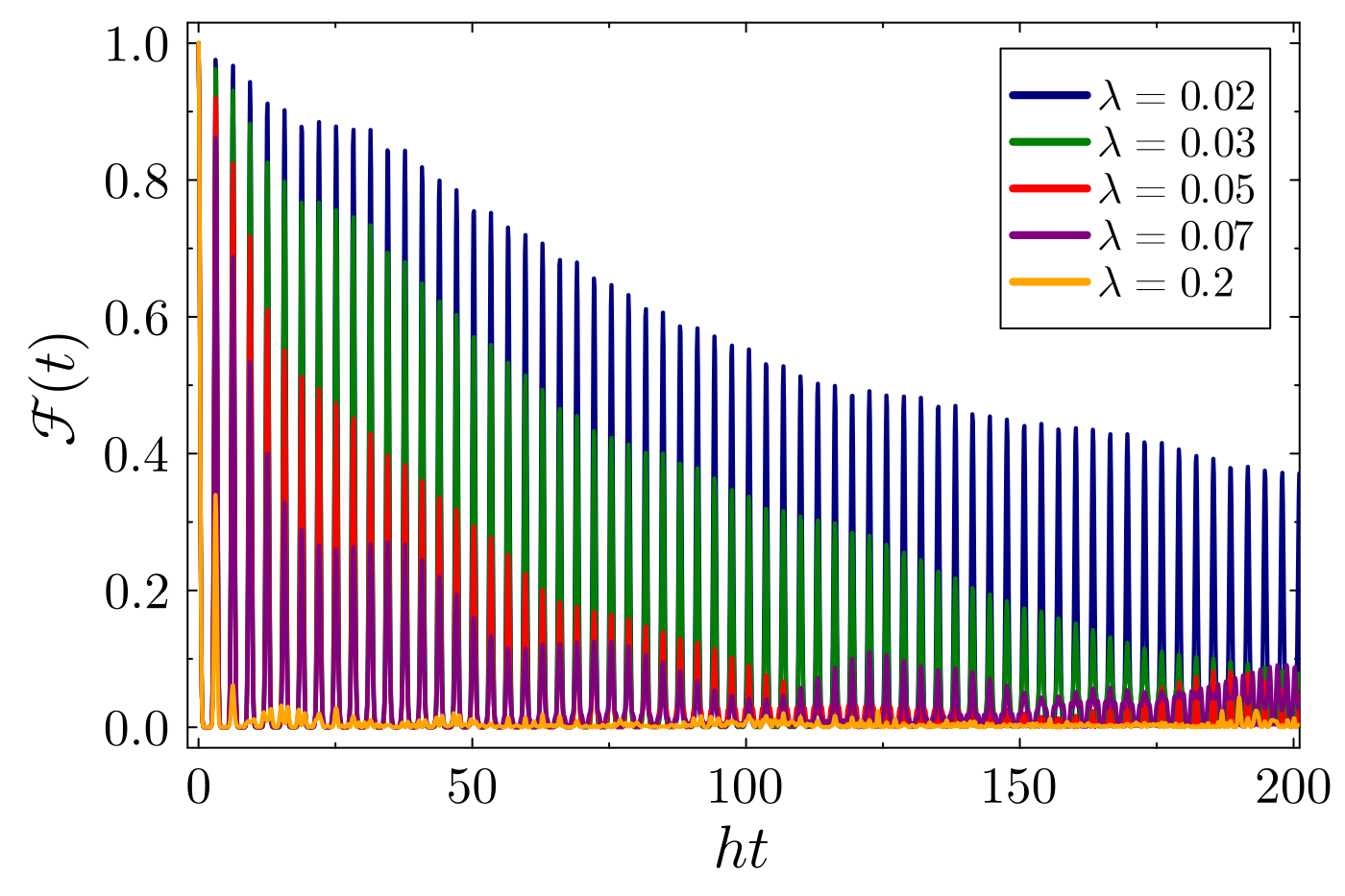}
        \put(21,59.7){\large (a)}
        \end{overpic}
        \phantomcaption\label{fig:ptb_fidelity}
    \end{subfigure}%
    \\[-4mm]
    \begin{subfigure}{\columnwidth}
        \centering
        \begin{overpic}[width=\columnwidth]     {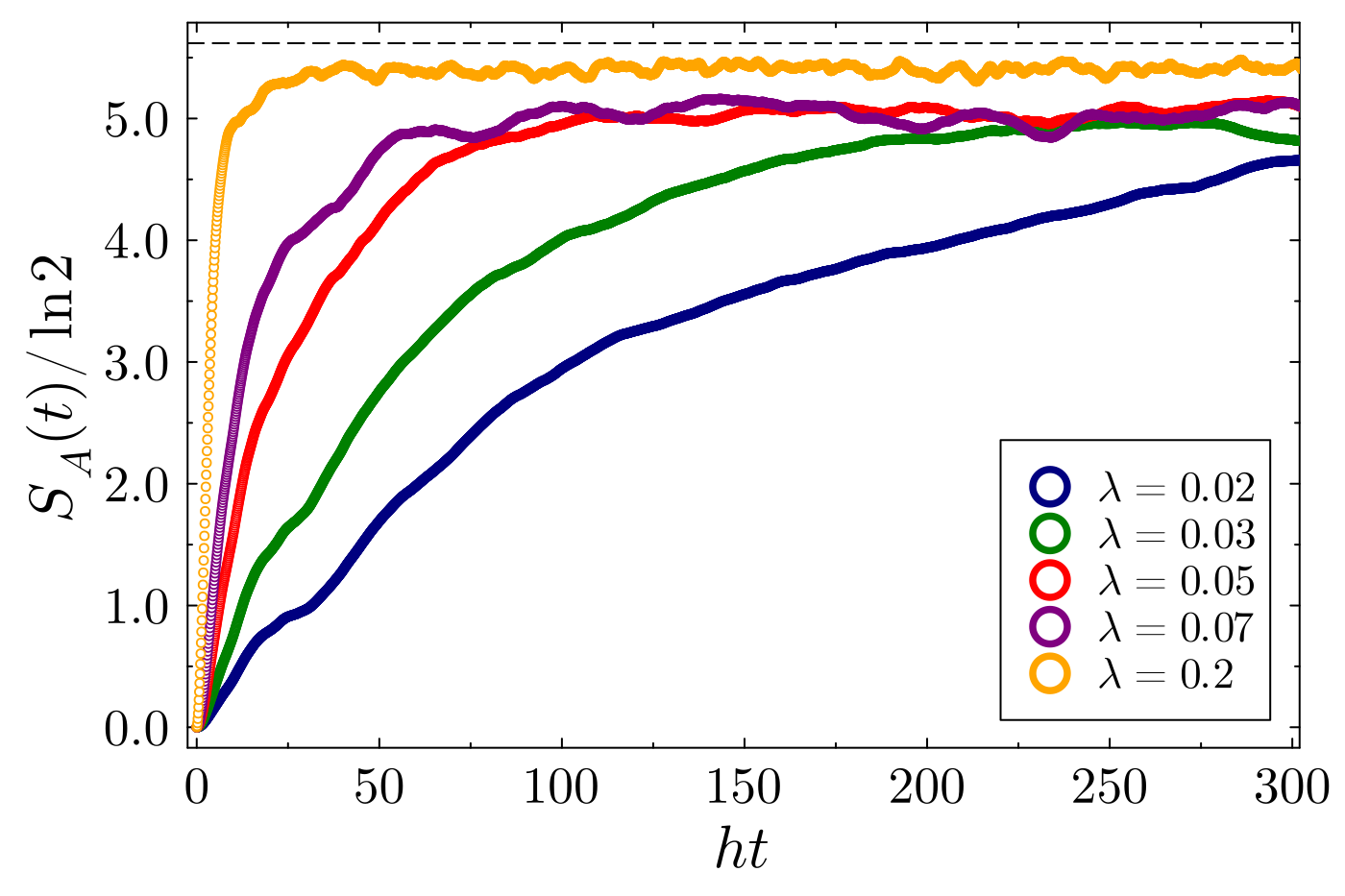}
        \put(21,59){\large (b)}
        \end{overpic}
        \phantomcaption\label{fig:ptb_entang_evol}
    \end{subfigure}
    \caption{Long-time evolution of an $L=8$ chain under the Hamiltonian $H$ Eq.~\eqref{H} after a quench from the nematic N\'eel state Eq.~\eqref{neel_state}, with rest of the parameters identical to those in Fig.~\ref{fig:unp_entang}: (a) Fidelity Eq.~\eqref{fidelity} and (b) bipartite entanglement entropy. Results are shown for increasing strength of perturbation Eqs.~\eqref{H} and \eqref{H_p} starting from $\lambda=0.02$.}
    \label{fig:ptb_fidelity_entang_evol}
\end{figure}

As presented in Fig.~\ref{fig:ptb_fidelity}, when the system is initialized in the N\'eel state, the many-body fidelity decays upon introducing a perturbation. For weak perturbations, persistent revivals survive for a finite duration, whereas strong perturbation causes the periodic oscillations to die down quickly. The entanglement dynamics of the same initial state, shown in Fig.~\ref{fig:ptb_entang_evol}, exhibit a gradual increase towards the entropy of a random state, with stronger perturbation leading to a faster growth. Conceptually, this implies that the initial N\'eel state no longer remains a product state under the Hamiltonian evolution in presence of the next nearest-neighbor exchange, which reflects the fact that the state can not be expressed as a superposition of special eigenstates of the full Hamiltonian $H$, unlike in the unperturbed model.

\begin{figure*}[t!]
    \centering 
    \begin{subfigure}{0.4\textwidth} 
        \centering 
        \begin{overpic}[width=\linewidth] {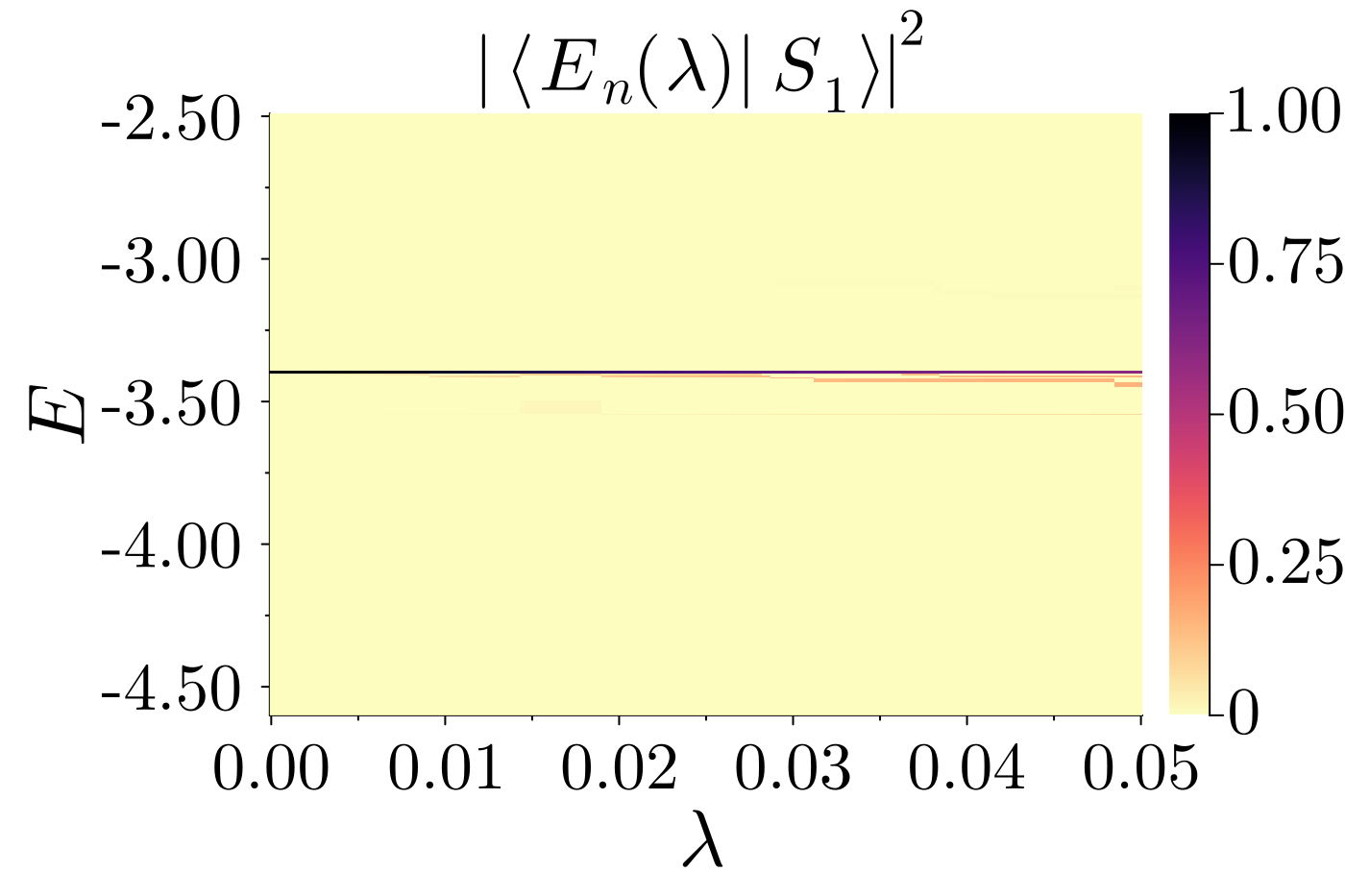} 
        \put(21,52.8){(a)} 
        \end{overpic} 
        \phantomcaption
        \label{fig:S1_overlap} 
    \end{subfigure} 
    \hspace{-0.7mm}
    \begin{subfigure}{0.4\textwidth} 
        \centering 
        \begin{overpic}[width=\linewidth] {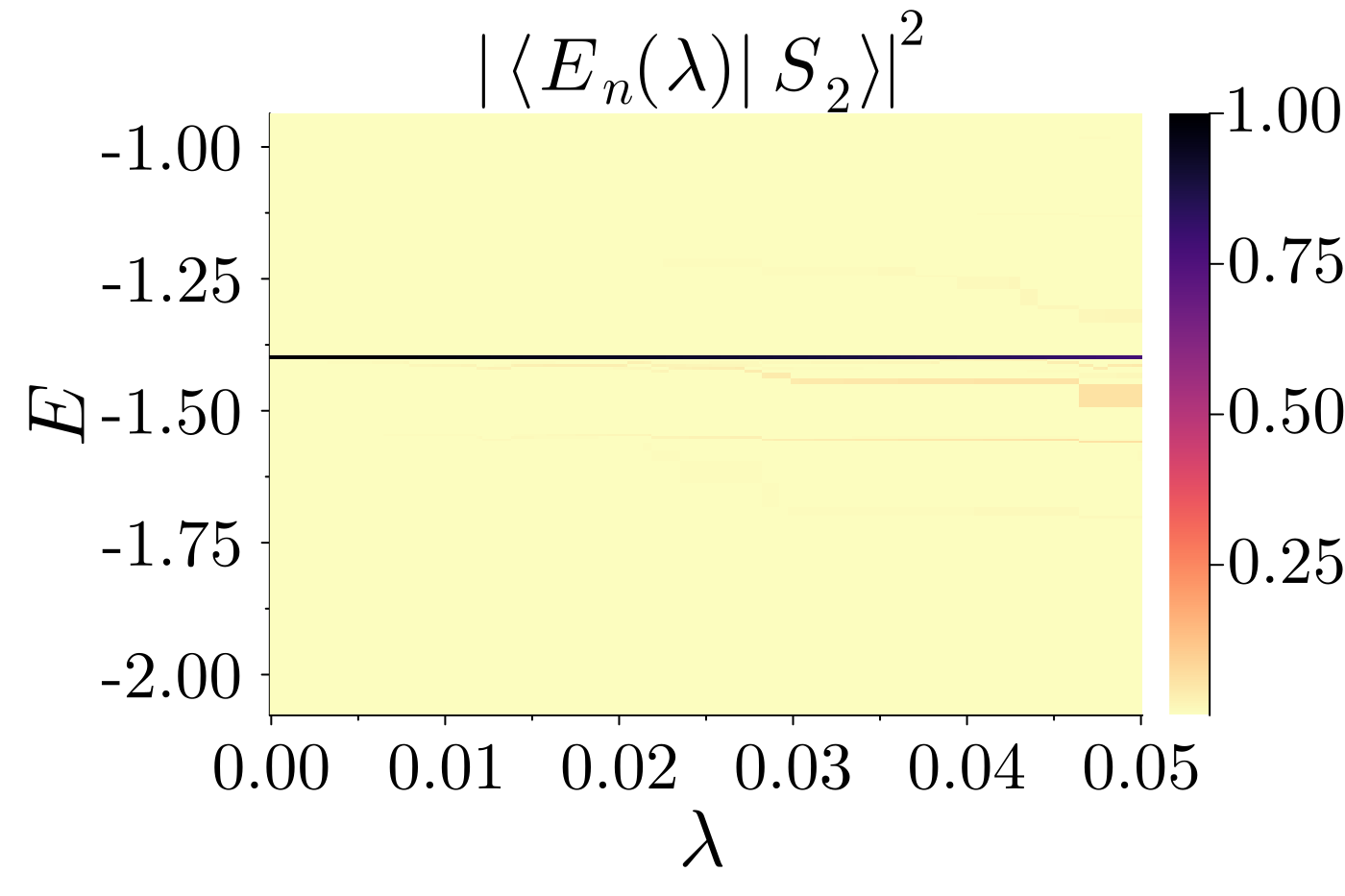} 
        \put(21,52.8){(b)} 
        \end{overpic} 
        \phantomcaption
        \label{fig:S2_overlap} 
    \end{subfigure} 
    \\[-4mm]
    \begin{subfigure}{0.4\textwidth} 
        \centering 
        \begin{overpic}[width=\linewidth] {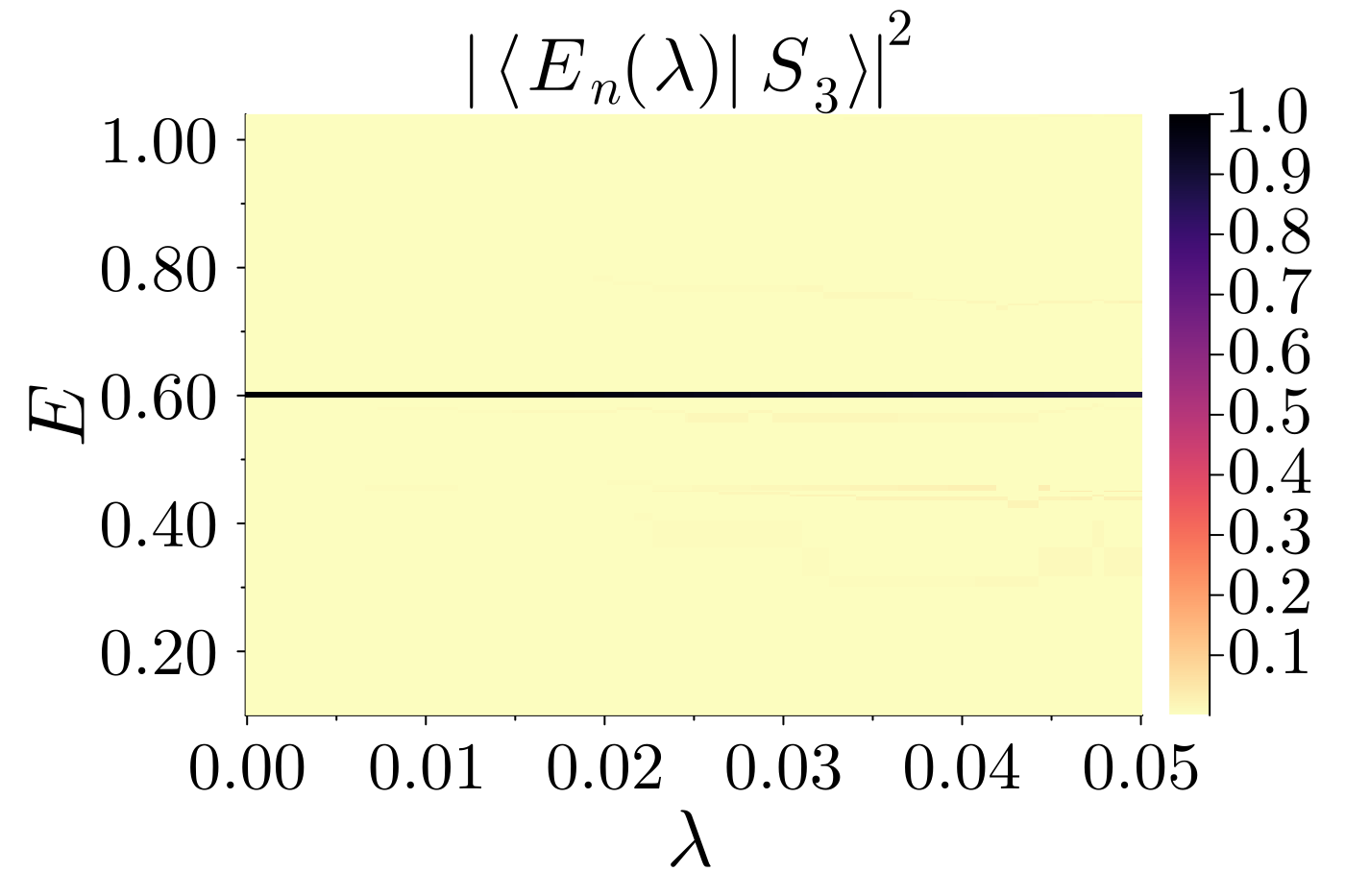} 
        \put(19.5,52.8){(c)} 
        \end{overpic} 
        \phantomcaption
        \label{fig:S3_overlap} 
    \end{subfigure} 
    \hspace{-0.7mm}
    \begin{subfigure}{0.4\textwidth} 
        \centering 
        \begin{overpic}[width=\linewidth] {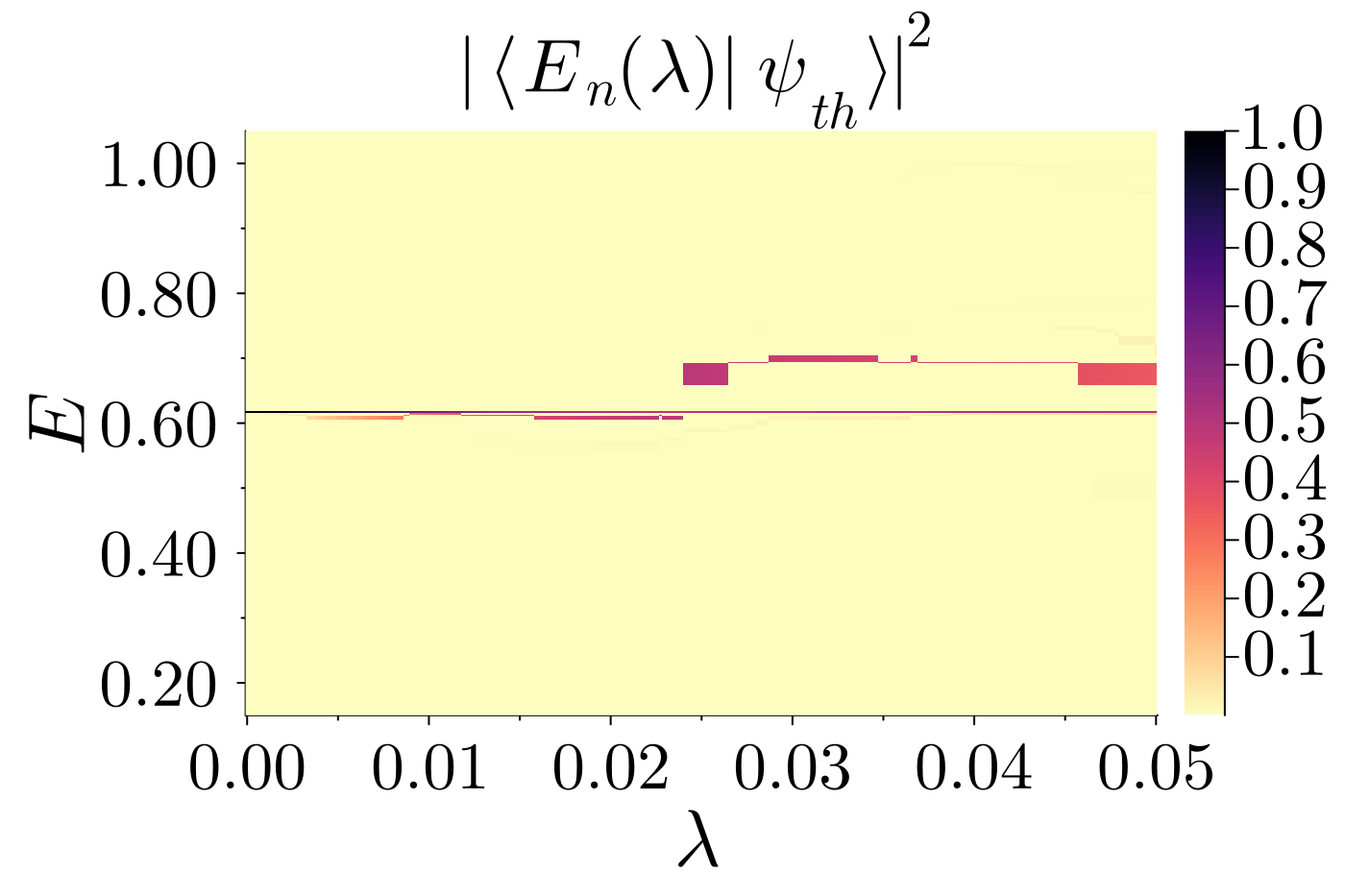} 
        \put(19.5,52.){(d)} 
        \end{overpic} 
        \phantomcaption
        \label{fig:th_overlap} 
    \end{subfigure} 
\caption{\justifying Heatmap plot of squared overlaps of the exact eigenstates $|E_n(\lambda)\rangle$ of the perturbed Hamiltonian $H_0+\lambda H_p$ for $L=6$, while the remaining parameters are same as in Fig.~\ref{fig:unp_entang}, with: the scarred states (a), (b), (c) the $|\mathcal{S}{}_1\rangle$, $|\mathcal{S}{}_2\rangle$, $|\mathcal{S}{}_3\rangle$ respectively, and (d) the thermal state $|\psi_{th}\rangle$. The thermal state is chosen as the eigenstate of the unperturbed Hamiltonian $H_0$ with eigenindex four more than the index of $|\mathcal{S}{}_3\rangle$ without any symmetry resolving. The plots simply show the projection of weights of the considered state on the perturbed manifold.}
\label{fig:scar_th_overlap} 
\end{figure*}

It is also important to examine the stability of these individual special eigenstates against perturbations that naturally arise in a system. In principle, the robustness of an eigenstate of $H_0$ is characterized by its ability to be deformed into an eigenstate of $H_0+\lambda H_p$ \cite{ryd_pert_silva}. The stability of the scarred states Eq.~\eqref{first_NN_scar} against the perturbation Eq.~\eqref{H_p} can be assessed examining the behavior of the fidelity as the perturbation strength $\lambda$ increases. Fig.~\ref{fig:scar_th_overlap} shows the squared overlaps of exact eigenstates of the perturbed Hamiltonian $H$ Eq.~\eqref{H} with the unperturbed scar states of $H_0$ Eq.~\eqref{H_0}, which is denoted as $|\langle E_n(\lambda)|\mathcal{S}{}_n\rangle|^2$, where $\ket{E_n(\lambda)}$ simply denotes the eigenstates of the Hamiltonian $H$, and $|\mathcal{S}{}_n\rangle$ are different scarred states under consideration.

As shown in Fig.~\ref{fig:S1_overlap}, $|\mathcal{S}{}_1\rangle$ maintains a dominant overlap with a single perturbed eigenstate up to $\lambda\approx 0.02$. Beyond this point, the overlap begins to spread across multiple eigenstates, which makes it difficult to pick out 
\noindent a unique descendant state of the unperturbed $|\mathcal{S}{}_1\rangle$ state.  Similarly, in Fig.~\ref{fig:S2_overlap}, sharp overlap of the two-bimagnon scarred state with a single eigenstate of the perturbed Hamiltonian $H$, i.e., $|\langle E_n(\lambda)|\mathcal{S}{}_2\rangle|^2$ indicates that there exists a single perturbed state which can be associated with the original scarred state till $\lambda\approx 0.02$. Once the perturbation strength increases, the state shows faint weights on additional eigenstates. In contrast, the state $|\mathcal{S}{}_3\rangle$ displays remarkable robustness in this small system size, as Fig.~\ref{fig:S3_overlap} vividly demonstrates no significant loss in fidelity throughout the entire range of perturbation strengths. Strikingly different behavior is observed for the unperturbed thermal state $|\psi_{th}\rangle$, chosen as a non-special eigenstate of the Hamiltonian $H_0$, which reveals strong hybridization with many perturbed eigenstates even at small values of $\lambda$, see Fig.~\ref{fig:th_overlap}. These observations suggest that the scarred eigenstates possess a certain degree of robustness against the specific perturbation introduced in Eq.~\ref{H_p}, compared to the generic eigenstates of the spin-1 \textit{XY} Hamiltonian $H_0$ for small chains. 

\subsection{Perturbation theory}
\label{sebsec:perturb_results}
Here we examine whether the scarred states remain accessible within perturbation framework for arbitrary system sizes. The validity of such a perturbative description up to a given order indicates that the characteristic features of a scar state remain stable to that order. Generic eigenstates residing in the middle of a dense spectrum are not expected to exhibit such stability. However, we show that the scarred states at finite energy density can be described perturbatively for small system sizes, while this estimation breaks down as the system size increases.

According to the nondegenerate perturbation theory, the energy of an unperturbed state $\ket{\mathcal{S}}$ can be perturbatively approximated up to the desired order as $E_{\mathrm{pert}}^{[N]}=E_{\mathcal{S}}^{(0)}+\sum_{n=1}^{N}\lambda^nE_{\mathcal{S}}^{(n)}$, where $E_{\mathcal{S}}^{(0)}$ is the unperturbed energy. The correction terms up to third order are given by,
\begin{figure*}[t!]
    \centering 
    \begin{subfigure}{0.325\textwidth} 
        \centering 
        \begin{overpic}[width=\linewidth] {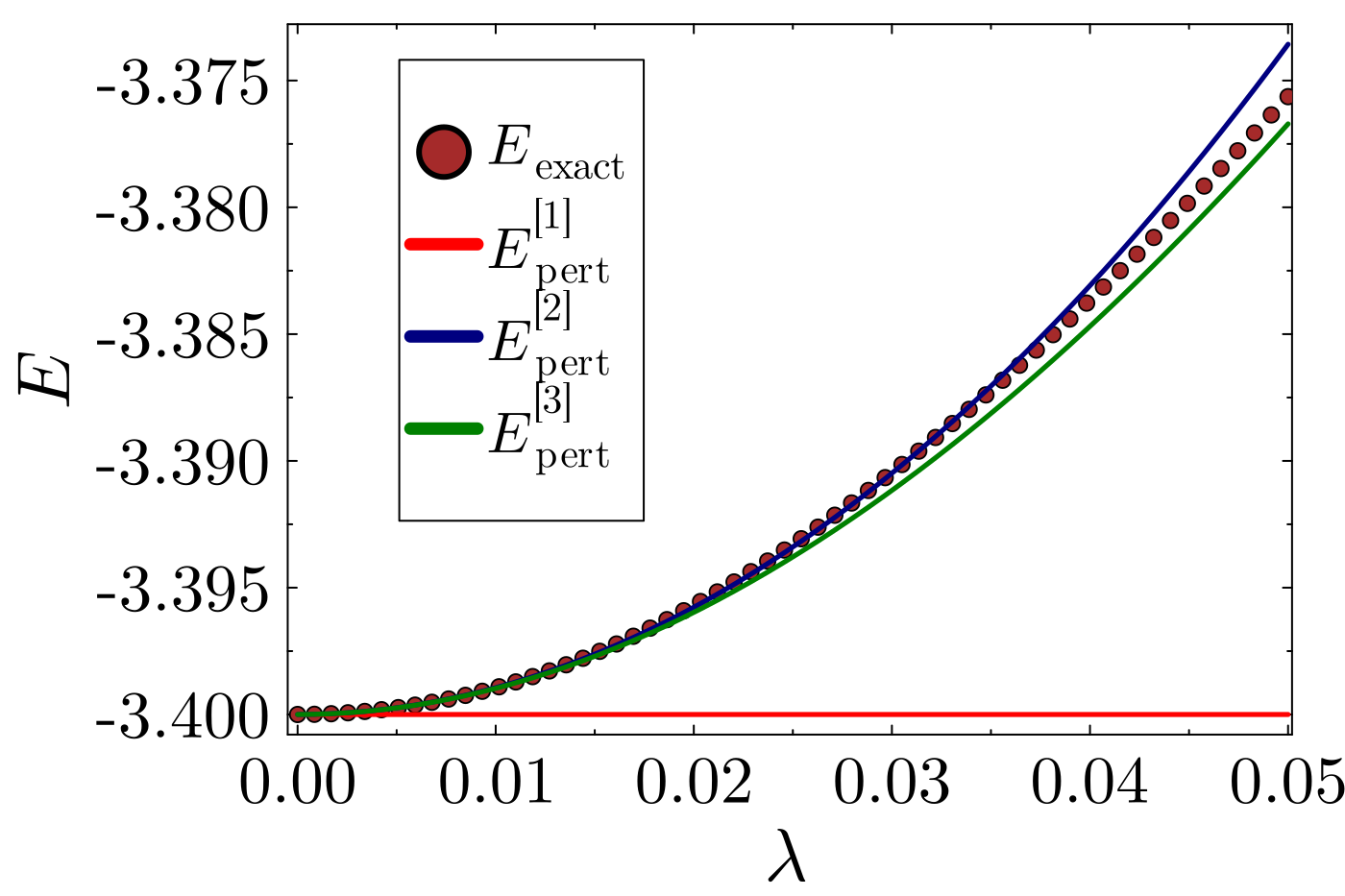} 
        \put(25,20){(a)} 
        \end{overpic} 
        \phantomcaption
        \label{fig:S1_pert_energy} 
    \end{subfigure} 
    \hspace{0.005mm}
    \begin{subfigure}{0.325\textwidth} 
        \centering 
        \begin{overpic}[width=\linewidth] {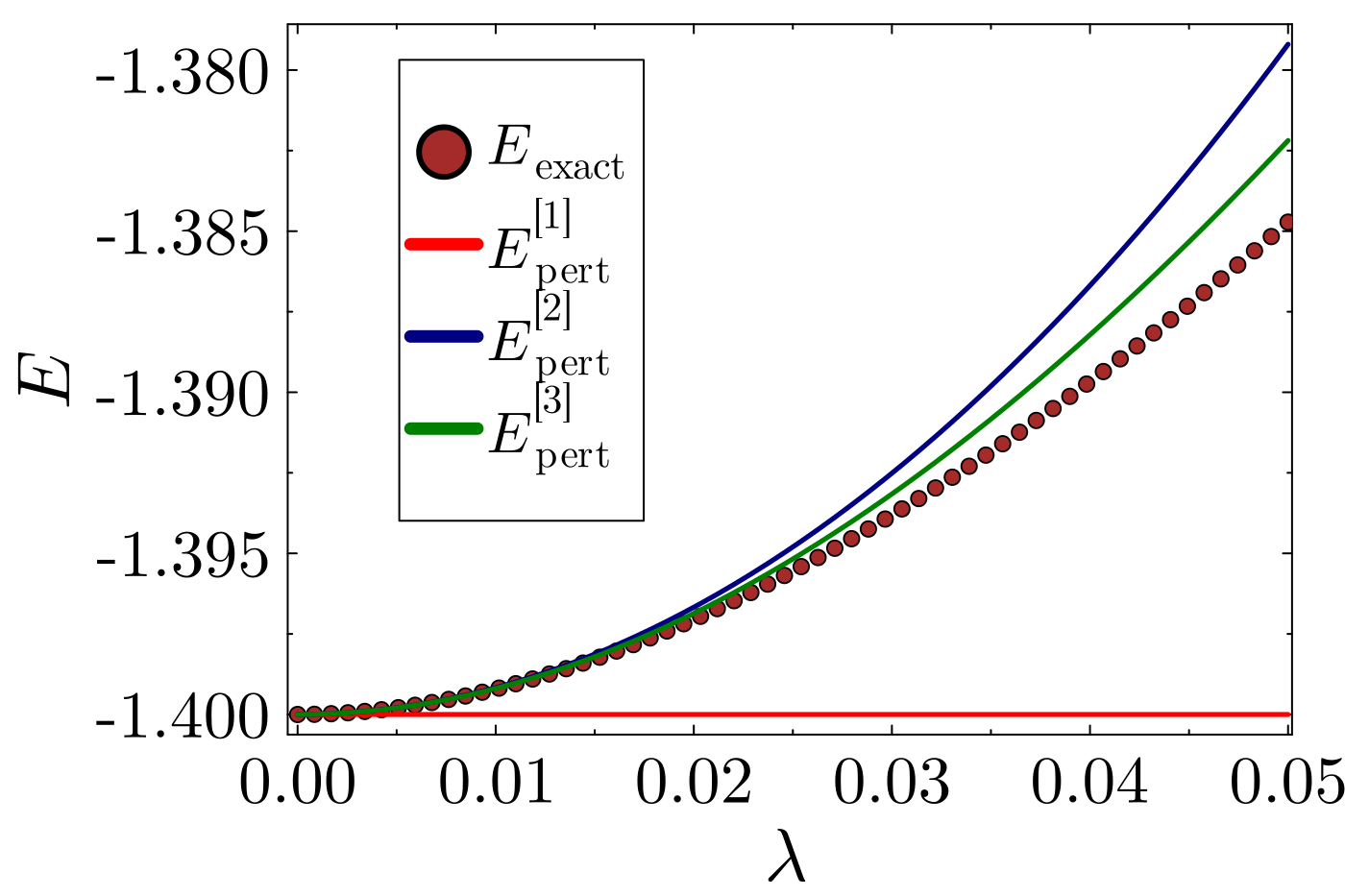} 
        \put(25,20){(b)} 
        \end{overpic} 
        \phantomcaption
        \label{fig:S2_pert_energy} 
    \end{subfigure} 
    \hspace{0.005mm}
    \begin{subfigure}{0.325\textwidth} 
        \centering 
        \begin{overpic}[width=\linewidth] {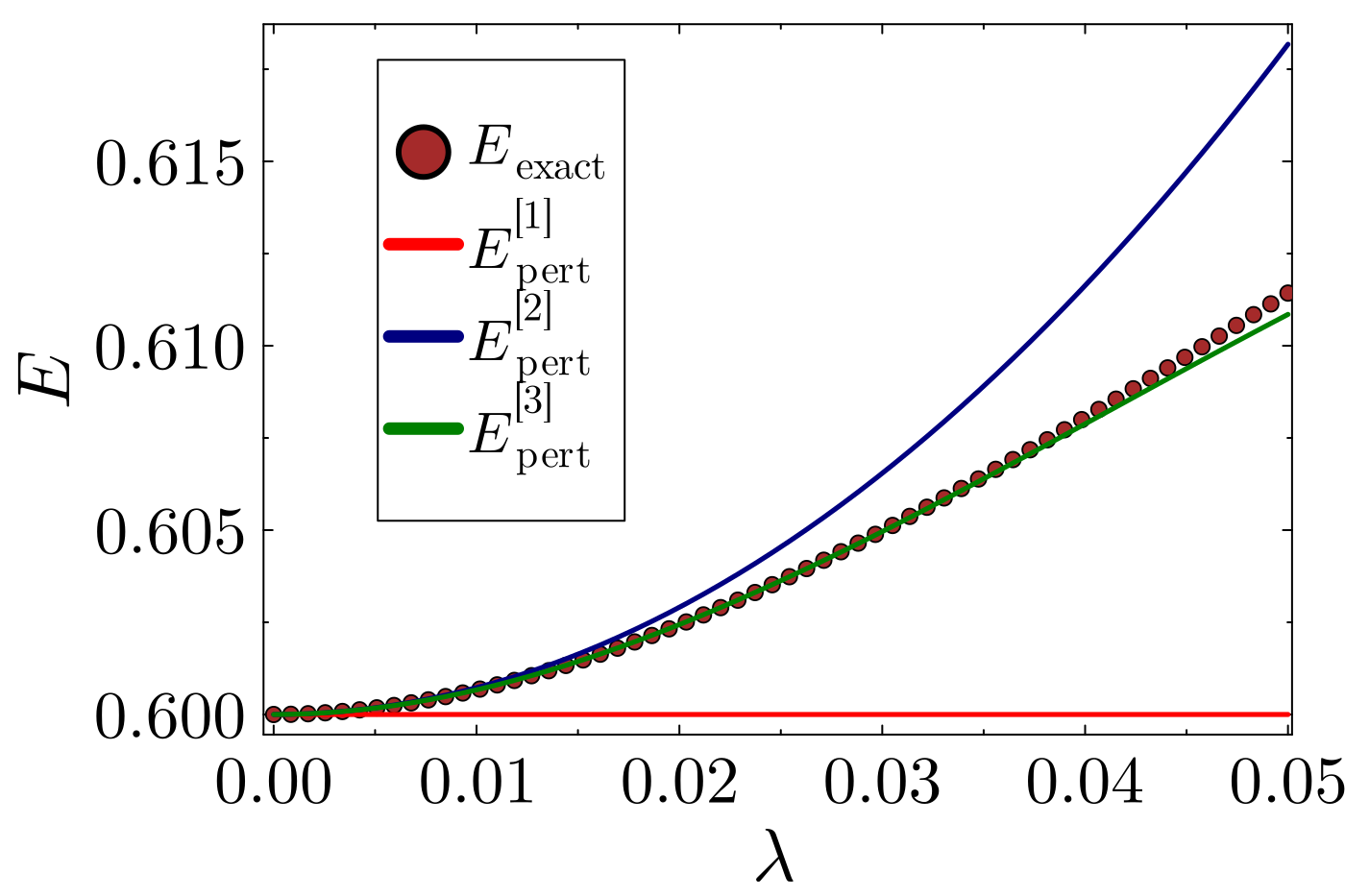} 
        \put(25,20){(c)} 
        \end{overpic} 
        \phantomcaption
        \label{fig:S3_pert_energy} 
    \end{subfigure} 
\caption{\justifying The perturbative energies $E_{\mathrm{pert}}$ (up to third order) and energies obtained from ED, $E_{\mathrm{exact}}$ are shown for different scarred states: (a), (b), and (c) the $|\mathcal{S}{}_1\rangle$, $|\mathcal{S}{}_2\rangle$, and $|\mathcal{S}{}_3\rangle$ respectively, for the same parameters as in Fig.~\ref{fig:scar_th_overlap}. $E_{\mathrm{exact}}$ corresponds to the energy of the eigenstate of the perturbed Hamiltonian $H$ Eq.~\eqref{H}, which has the largest overlap with the scarred state under consideration. The first-order correction is zero, as expected, because the perturbation $H_p$ is completely off-diagonal in the working $S_z$ basis. For small perturbation strengths $\lambda$, the third-order correction provides a reliable approximation to the ED results. The closest agreement between the perturbation theory and exact diagonalization is observed for the state $|\mathcal{S}{}_3\rangle$.}
\label{fig:pert_energy} 
\end{figure*} 
\begin{align}
    E_{\mathcal{S}}^{(1)} &= \bra{\mathcal{S}}H_p\ket{\mathcal{S}}, \\
    E_{\mathcal{S}}^{(2)} &= \sum_{m\neq \mathcal{S}}\frac{|\bra{m^{(0)}}H_p\ket{\mathcal{S}}|^2}{E_{\mathcal{S}}^{(0)}-E_m^{(0)}}, \\
    E_{\mathcal{S}}^{(3)} &= \sum_{m\neq \mathcal{S}} \sum_{n\neq \mathcal{S}} \frac{\bra{\mathcal{S}}H_p\ket{m^{(0)}} \bra{m^{(0)}}H_p\ket{n^{(0)}} 
    \bra{n^{(0)}}H_p\ket{\mathcal{S}}}{\left(E_{\mathcal{S}}^{(0)}-E_{m}^{(0)}\right)
    \left( E_{\mathcal{S}}^{(0)}-E_{n}^{(0)}\right)} \nonumber 
\end{align}
\begin{align}
    - \bra{\mathcal{S}}H_p\ket{\mathcal{S}} 
    \sum_{m\neq \mathcal{S}} 
    \frac{
    |\bra{m^{(0)}}H_p\ket{\mathcal{S}}|^2
    }{
    \left(E_{\mathcal{S}}^{(0)}-E_{m}^{(0)}\right)^2
    }.
\end{align}
Note that, in this section, the unperturbed scarred states are denoted by $\ket{\mathcal{S}}$ instead of $\ket{\mathcal{S}{}_n}$ in certain places. 

Fig.~\ref{fig:pert_energy} presents perturbative corrections to energy for three scarred states $\ket{\mathcal{S}{}_1}$, $\ket{\mathcal{S}{}_2}$, and $\ket{\mathcal{S}{}_3}$, along with their corresponding ED energies $E_{\mathrm{exact}}$. $E_{\mathrm{exact}}$ represents the energy of the eigenstate which is identified as the descendant state of the scarred state through the largest overlap. As seen in the figures, third-order perturbation theory captures the energy behavior well for small perturbation strengths. Remarkably, for the state $\ket{\mathcal{S}{}_3}$, third order correction yields an excellent agreement between $E_{\mathrm{pert}}^{[3]}$ and $E_{\mathrm{exact}}$ even at larger values of $\lambda$. The strong nature of the perturbation is evident from the fact that first-order correction completely fails to reproduce the energy, as the perturbation results in mixing between different levels. 

Now we investigate the agreement between the wave function of a scarred state obtained from exact diagonalization and the corresponding perturbed wave function. In the Rayleigh-Schrödinger perturbation framework, the perturbed wave function (without normalization) to second order can be expressed as $\ket{\mathcal{S}_{\mathrm{pert}}}=\ket{\mathcal{S}}+\lambda \ket{\mathcal{S}^{(1)}}+\lambda^2\ket{\mathcal{S}^{(2)}}$, where
\begin{align}
    \ket{\mathcal{S}^{(1)}} &= \sum_{m\neq \mathcal{S}} \frac{\bra{m^{(0)}}H_p\ket{\mathcal{S}}}{E_\mathcal{S}^{(0)}-E_m^{(0)}}, \\
    \label{1st_pert_wf}
    \ket{\mathcal{S}^{(2)}} &= \sum_{m\neq \mathcal{S}}\Biggl[\sum_{n\neq \mathcal{S}}\frac{\bra{m^{(0)}}H_p\ket{n^{(0)}}\bra{n^{(0)}}H_p\ket{\mathcal{S}}}{\left(E_{\mathcal{S}}^{(0)} -E_{m}^{(0)}\right)\left(E_{\mathcal{S}}^{(0)}-E_{n}^{(0)}\right)} \nonumber\\ 
    &\quad-\frac{\bra{m^{(0)}}H_p\ket{\mathcal{S}}\bra{\mathcal{S}}H_p\ket{\mathcal{S}}}{\left(E_{\mathcal{S}}^{(0)}-E_{m}^{(0)}\right)^2}\Biggr]
\end{align}
\noindent To assess the validity of this perturbatively constructed wave function, we numerically evaluate $F_{\mathrm{pert}}(\lambda;\mathcal{S})=\frac{|\braket{\mathcal{S}|\mathcal{S}_{\mathrm{pert}}}|^2}{\braket{\mathcal{S}_{\mathrm{pert}}|\mathcal{S}_{\mathrm{pert}}}}$ and contrast it with $F_{\mathrm{exact}}(\lambda;\mathcal{S})=|\braket{E_n(\lambda)|\mathcal{S}}|^2$, where $\ket{E_n(\lambda)}$ denotes the descendant state of $\ket{\mathcal{S}{}_n}$ obtained from ED.

\begin{figure*}[t!]
    \centering 
    \begin{subfigure}{0.325\textwidth} 
        \centering 
        \begin{overpic}[width=\linewidth] {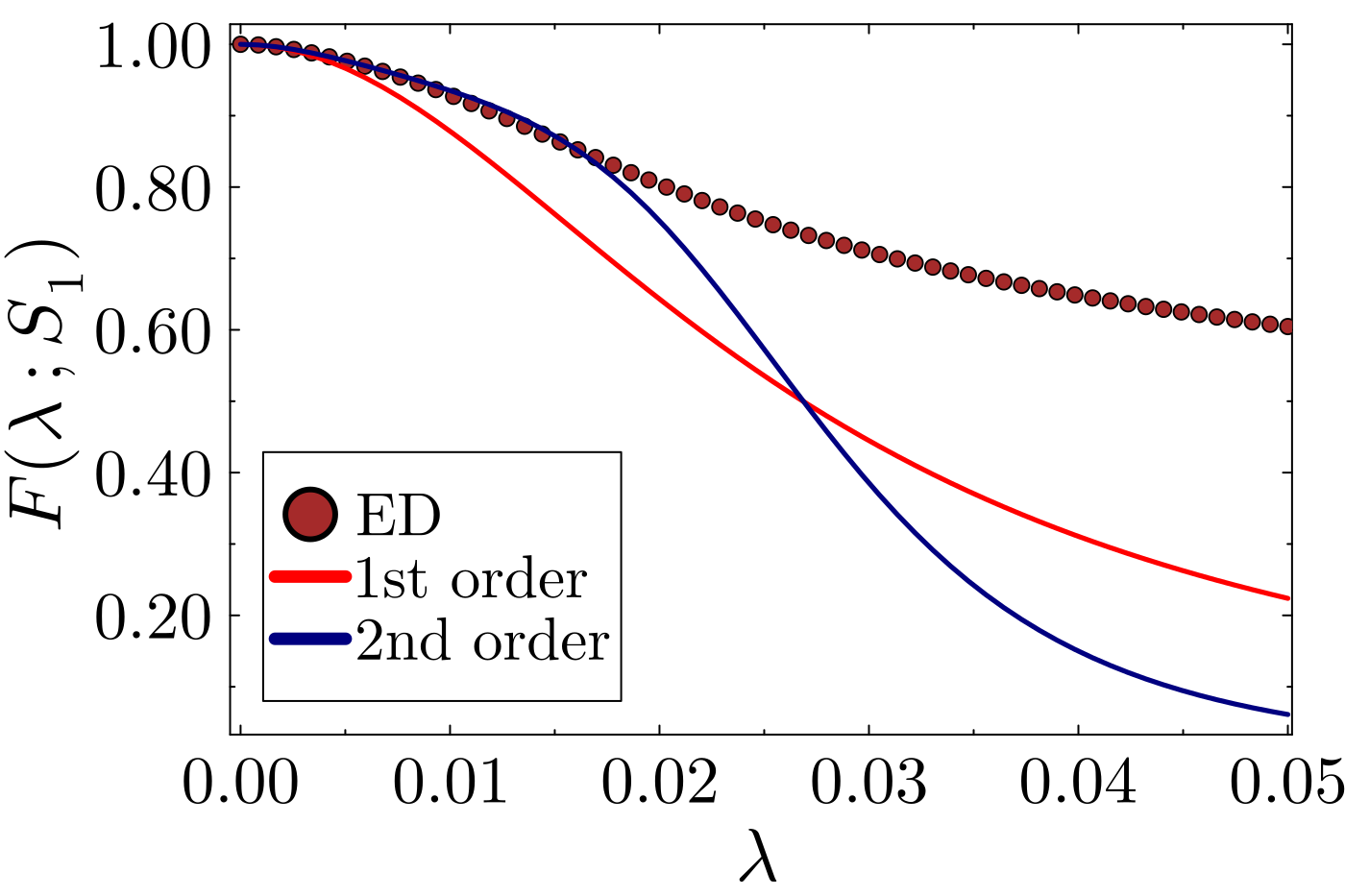} 
        \put(19,54){(a)} 
        \end{overpic} 
        \phantomcaption
        \label{fig:S1_pert_overlap} 
    \end{subfigure} 
    \hspace{0.005mm}
    \begin{subfigure}{0.325\textwidth} 
        \centering 
        \begin{overpic}[width=\linewidth] {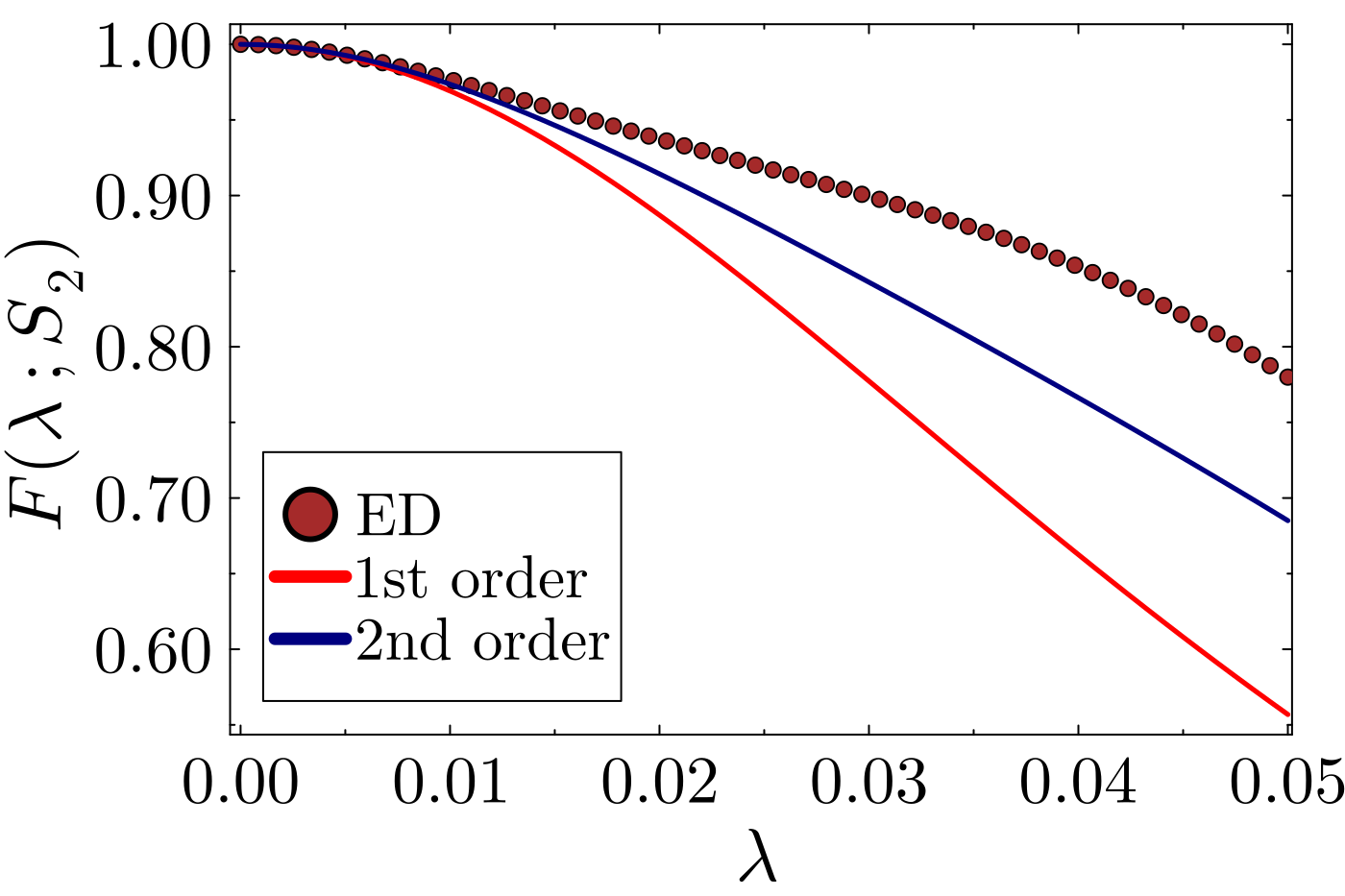} 
        \put(19,54){(b)} 
        \end{overpic} 
        \phantomcaption
        \label{fig:S2_pert_overlap} 
    \end{subfigure} 
    \hspace{0.005mm}
    \begin{subfigure}{0.325\textwidth} 
        \centering 
        \begin{overpic}[width=\linewidth] {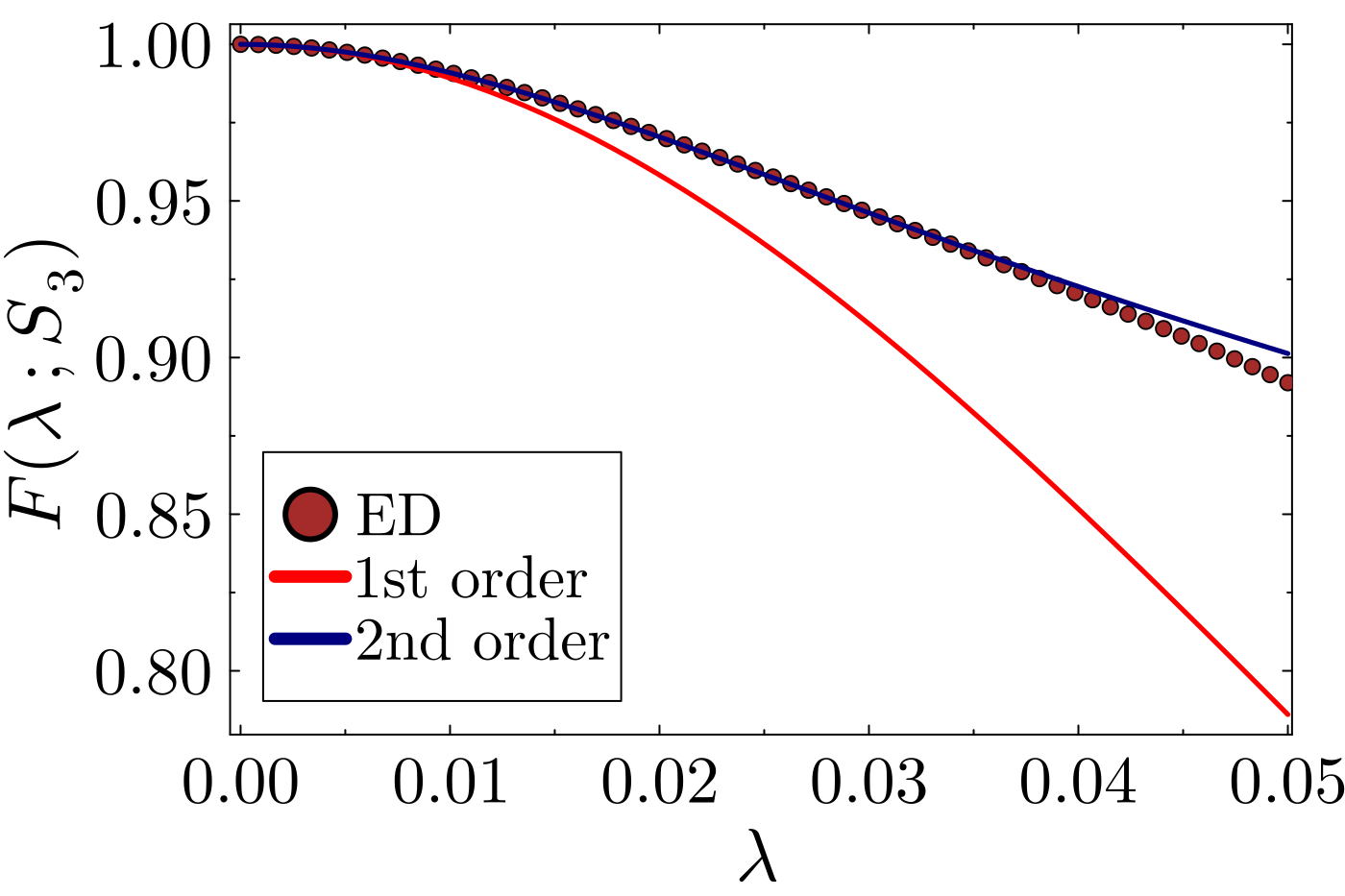} 
        \put(19,54){(c)} 
        \end{overpic} 
        \phantomcaption
        \label{fig:S3_pert_overlap} 
    \end{subfigure} 
\caption{\justifying Comparison of the overlaps $F(\lambda;\mathcal{S})$ obtained from perturbatively constructed wave functions (up to second order) and ED wave functions for (a) $|\mathcal{S}{}_1\rangle$, (b) $|\mathcal{S}{}_2\rangle$, and (c) $|\mathcal{S}{}_3\rangle$. The set of parameters is the same as in Fig.~\ref{fig:scar_th_overlap}. In small $\lambda$ regime, second-order perturbative analysis captures the behavior of the scarred eigenstates well.}
\label{fig:pert_overlap} 
\end{figure*}

As demonstrated in Fig.~\ref{fig:pert_overlap}, second-order correction yields a reasonably accurate approximation of the perturbed wave function of the scarred states when perturbation strength is small, whereas even the first-order wavefunction captures the ED results well in the extremely weak perturbation limit. The numerical values of the ED overlaps for different scarred states, presented in Figs.~\ref{fig:scar_th_overlap}\hyperref[fig:S1_overlap]{(a)}–\hyperref[fig:S3_overlap]{(c)}, are more clearly shown in Figs.~\ref{fig:pert_overlap}\hyperref[fig:S1_pert_overlap]{(a)}–\hyperref[fig:S3_pert_overlap]{(c)}. The best agreement between the perturbative and ED overlaps is observed for the three-bimagnon state $\ket{\mathcal{S}{}_3}$. This behavior originates from the fact that $\ket{\mathcal{S}{}_3}$ remains particularly robust under weak perturbations at this small system size because of negligible overlaps with other eigenstates. For $\ket{\mathcal{S}{}_1}$ and $\ket{\mathcal{S}{}_2}$, second-order perturbative overlap $F_{\mathrm{pert}}$ closely follows $F_{\mathrm{exact}}$ up to $\lambda\approx 0.02$. Beyond this point, the perturbative estimate begins to deviate from the ED result as the weights of the scarred state become significantly distributed over multiple states at nearby energies. These observations suggest that the perturbative analysis remains reliable in the weak perturbation limit for this small system size. 

\subsection{Signatures of thermalization}
\label{subsec:thermalization}
Despite the success of perturbation theory in describing the scarred states in presence of weak next nearest-neighbor exchange in small chains, we show that signatures of thermalization emerge in the late time dynamics of these special eigenstates and in the finite-size scaling of the perturbation matrix elements that couple the scar states to other eigenstates of the \textit{XY} Hamiltonian. 

The eigenstates $\ket{n^{(0)}}$ of a thermalizing system effectively act as random vectors in the Hilbert space. When such a system is subjected to a perturbation $H_p$, according to the random matrix theory \cite{rand_mat_mehta}, the off-diagonal 
{\setlength{\belowcaptionskip}{0pt}
\begin{figure}[H]
    \centering 
    \begin{subfigure}{0.487\columnwidth} 
        \centering 
        \begin{overpic}[width=\linewidth] {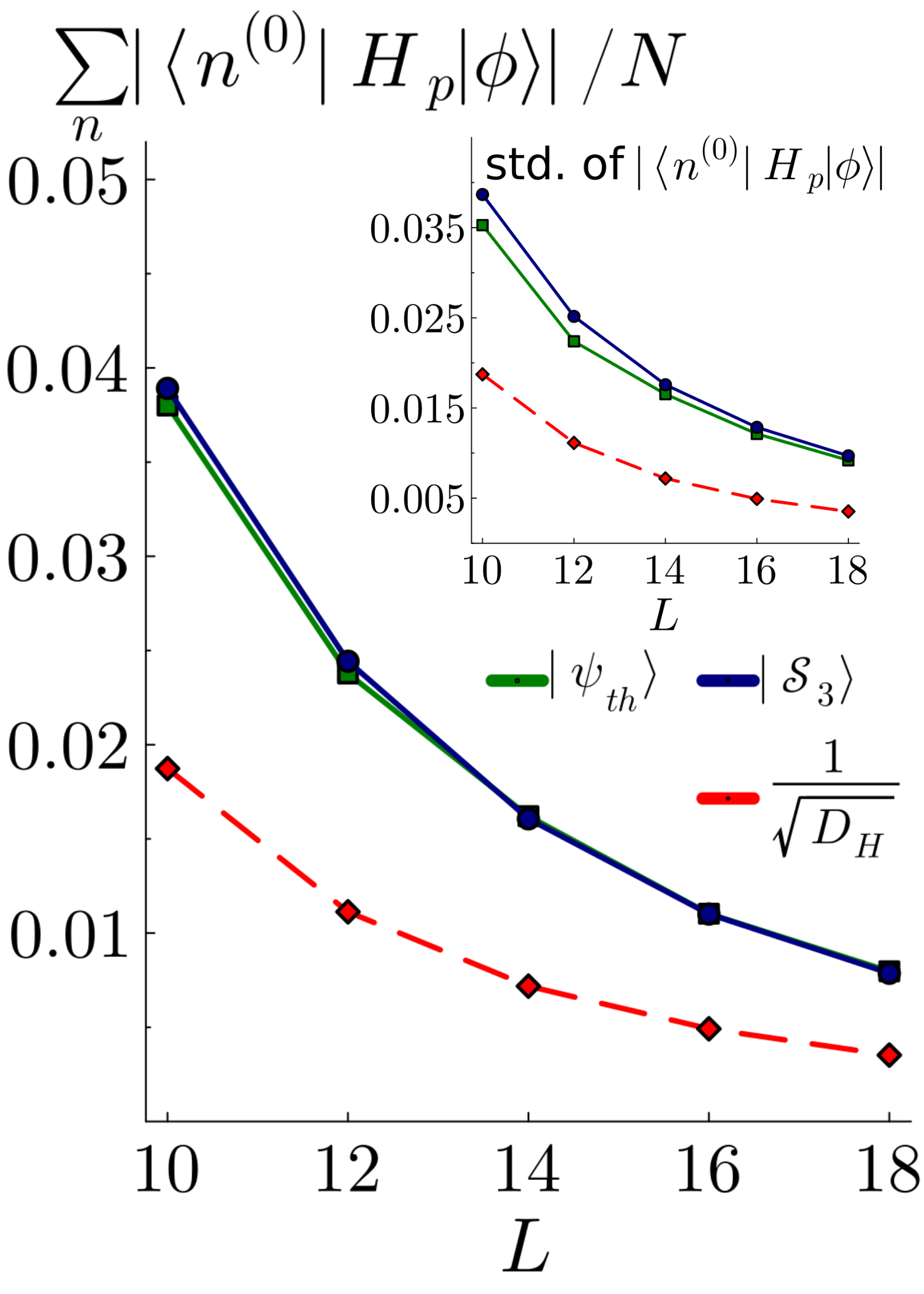} 
        \put(15,18){(a)} 
        \end{overpic} 
        \phantomcaption
        \label{fig:S3_pert_scaling_ediff} 
    \end{subfigure} 
    \hspace{0.00001mm}
    \begin{subfigure}{0.487\columnwidth} 
        \centering 
        \begin{overpic}[width=\linewidth] {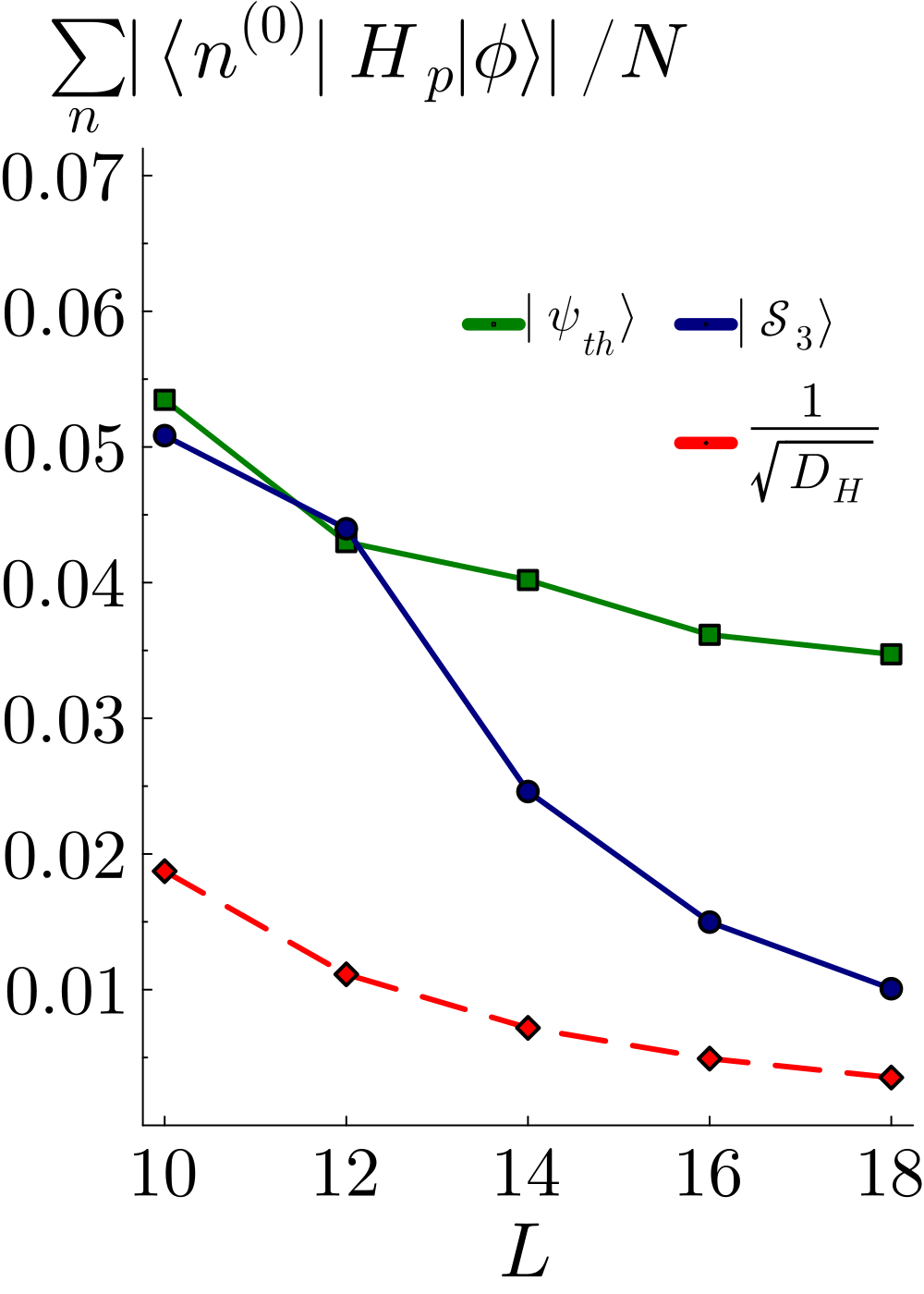} 
        \put(15,18){(b)} 
        \end{overpic} 
        \phantomcaption
        \label{fig:S3_pert_scaling_smallest} 
    \end{subfigure} 
    \caption{Finite-size scaling of the matrix elements: (a) $|\langle n^{(0)}|H_p|\phi\rangle|$ averaged over the eigenstates $\ket{n^{(0)}}$ satisfying $|E_n^{(0)}-E_{\phi}^{(0)}|<3$. Inset shows scaling of standard deviation of $|\langle n^{(0)}|H_p|\phi\rangle|$ calculated within the same energy window. (b) Average of $|\langle n^{(0)}|H_p|\phi\rangle|$ over $800$ eigenstates having the smallest $|E_n^{(0)}-E_{\phi}^{(0)}|$. $\ket{\phi}$ corresponds to $\ket{\mathcal{S}{}_3}$ for the blue line and to the thermal state $\ket{\psi_{th}}$ for the green line. $\ket{\psi_{th}}$ is the eigenstate of $H_0$ with eigenindex four more than that of $\ket{\mathcal{S}{}_3}$ in the same $S^z_{\mathrm{tot}}$ sector.}
    \label{fig:S3_pert_element_scaling}
\end{figure}}
\noindent matrix elements $\langle n^{(0)}|H_p|\psi\rangle$ between a state $\ket{\psi}$ and eigenstates $\ket{n^{(0)}}$ in nearby energies vary as $D_H^{-1/2}$ \cite{ryd_pert_chandran}, where $D_H$ is the Hilbert space dimension. This scaling follows from the fact that the overlap of a random vector with another vector pointing to some fixed direction in the Hilbert space, such as $H_p\ket{\psi}$, typically scales in the order of $D_H^{-1/2}$. Remarkably, this relation holds true for any $\ket{\psi}$ that is not an eigenstate of $H_p$, irrespective of whether $\ket{\psi}$ itself is thermal or nonthermal.

For our system, the matrix element $|\langle n^{(0)}|H_p|\psi_{th}\rangle|$ for the thermal state $\ket{\psi_{th}}$ approximately scales as $1/\sqrt{D_H}$ when averaged over the energy window $|E_n^{(0)}-E_{\psi_{th}}^{(0)}|<3$, as shown in Fig.~\ref{fig:S3_pert_scaling_ediff}. It is important to note that the perturbation $H_p$ preserves the U(1) symmetry and therefore couples only states within the same $S^z_{\mathrm{tot}}$ sector. Hence, $D_H$ is the Hilbert space dimension of the total magnetization sector in which $\ket{\psi_{th}}$ resides. To rule out arbitrariness in the averaging protocol, result for $800$ eigenstates with the smallest energy difference is shown in Fig.~\ref{fig:S3_pert_scaling_smallest}. Notably, for both averaging procedures, the averaged off-diagonal matrix element for the scarred state $\ket{\mathcal{S}{}_3}$ also exhibits approximately the same $D_H^{-1/2}$ scaling, apart from subleading deviations. It can be seen that the numerical values of the averaged matrix elements for both the thermal state and the scar state are also nearly the same in Fig.~\ref{fig:S3_pert_scaling_ediff} (for the distribution of perturbation matrix elements in the spectrum, see Fig.~\ref{fig:pert_matrix_element_dist} in Appendix \ref{appendix:dist_pert_matrix_elem}). Moreover, the density of states at infinite-temperature energy density increases as $D_H$, which implies that the ratio of the perturbation matrix elements to the energy-level spacing in Eq.~\eqref{1st_pert_wf} will scale as $\sqrt{D_H}$. This indicates that the scar states hybridize strongly with ETH-obeying eigenstates at nearby energies and will ultimately thermalize as the system size increases. A similar conclusion can also be drawn for $\ket{\mathcal{S}{}_2}$, based on its scaling behavior (see Fig.~\ref{fig:S2_pert_element_scaling} in Appendix \ref{appendix:add_overlap_scaling}).

As discussed earlier in Sec.~\ref{sec:model}, the nonthermal nature of many-body scars is manifested through persistent oscillations in unitary dynamics following a quantum quench from an initial state with large overlaps with the scarred states. In contrast, Fig.~\ref{fig:unp_fidelity} reveals a rapid relaxation of fidelity when the system is initialized in a generic thermalizing state. As shown in Fig.~\ref{fig:ptb_fidelity}, introduction of the perturbation leads to damping of the fidelity amplitude of the nematic state Eq.~\eqref{neel_state} which behaves as a nongeneric, thermalization-evading initial state for the unperturbed \textit{XY} chain. This suggests that the scarred eigenstates 
\begin{figure}[H]
    \centering
    \begin{subfigure}{\columnwidth}
        \centering
        \begin{overpic}[width=\columnwidth]       {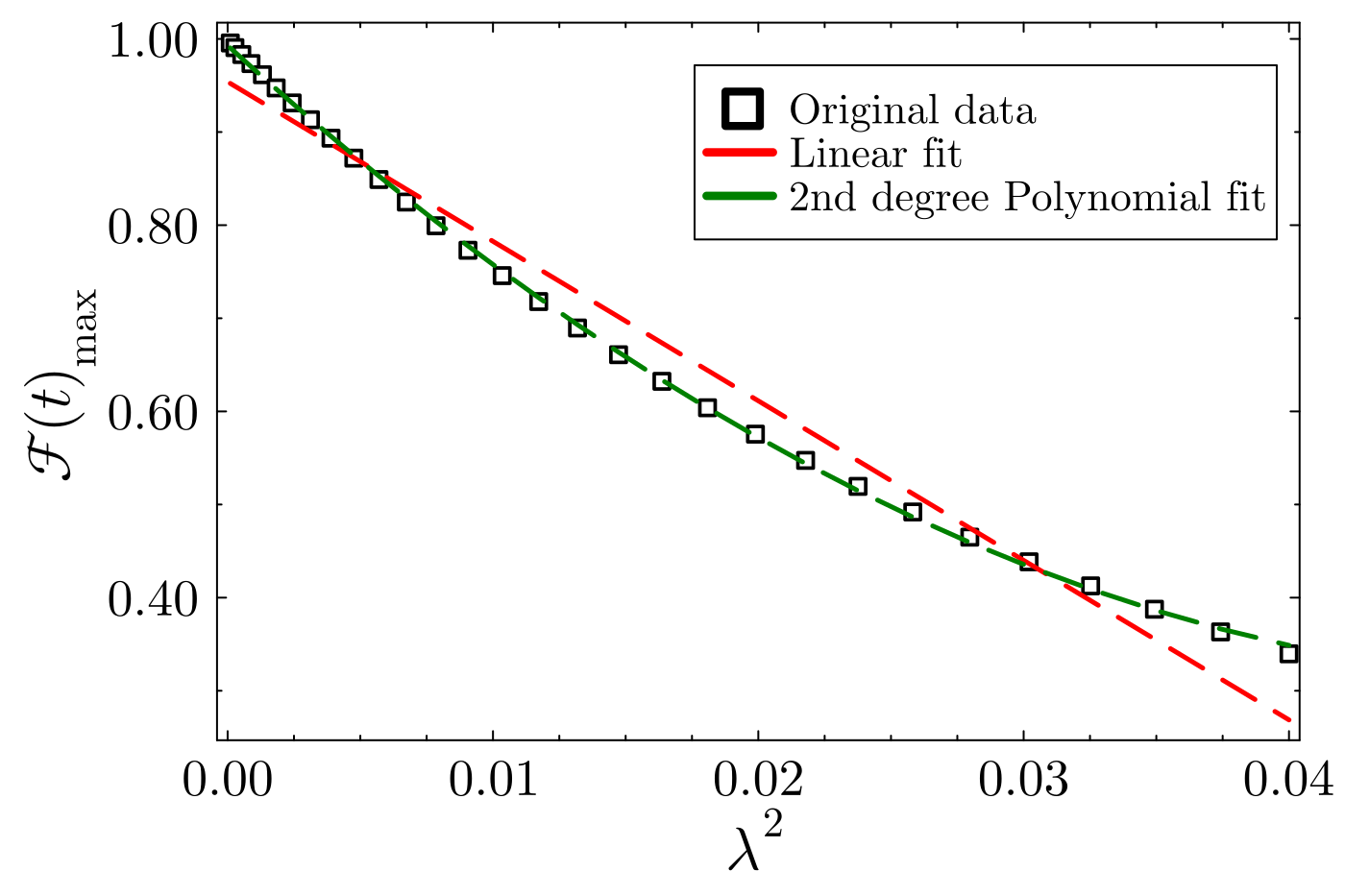}
        \put(18,15){\large (a)}
        \end{overpic}
       \phantomcaption\label{fig:fidelity_decay_fit}
    \end{subfigure}%
    \\[-4mm]
    \begin{subfigure}{\columnwidth}
        \centering
        \begin{overpic}[width=\columnwidth]     {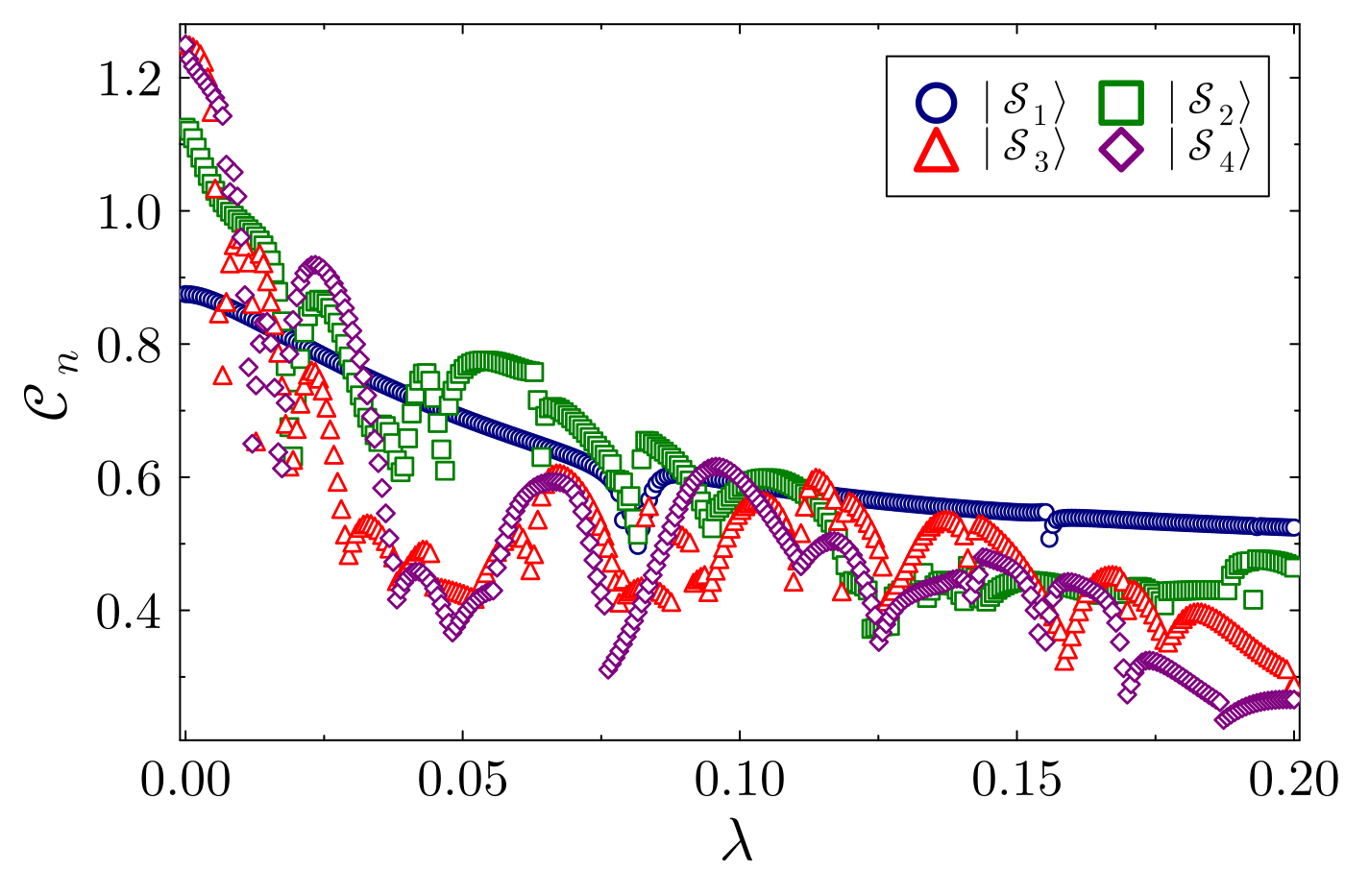}
        \put(16,15){\large (b)}
        \end{overpic}
        \phantomcaption\label{fig:ptb_odlro}
    \end{subfigure}
    \caption{(a) Dependence of the first peak of fidelity in Fig.~\ref{fig:ptb_fidelity} on the squared perturbation strength. The maximum height $\mathcal{F}_{\mathrm{max}}(t)$ varies with $\lambda^2$ as $0.9929-26.0004\,\lambda^2 +247.4208\,(\lambda^2)^2$. (b) Decay of the off-diagonal long-range order (ODLRO) Eq.~\eqref{odlro} for different scarred states with increasing perturbation strength $\lambda$. Same parameters of Fig.~\ref{fig:unp_fidelity} have been used in (a) and (b).}
    \label{fig:fidelity_decay_fit_odlro}
\end{figure}
\begin{figure}[H]
    \centering
    \begin{subfigure}{\columnwidth}
        \centering
        \begin{overpic}[width=\columnwidth]       {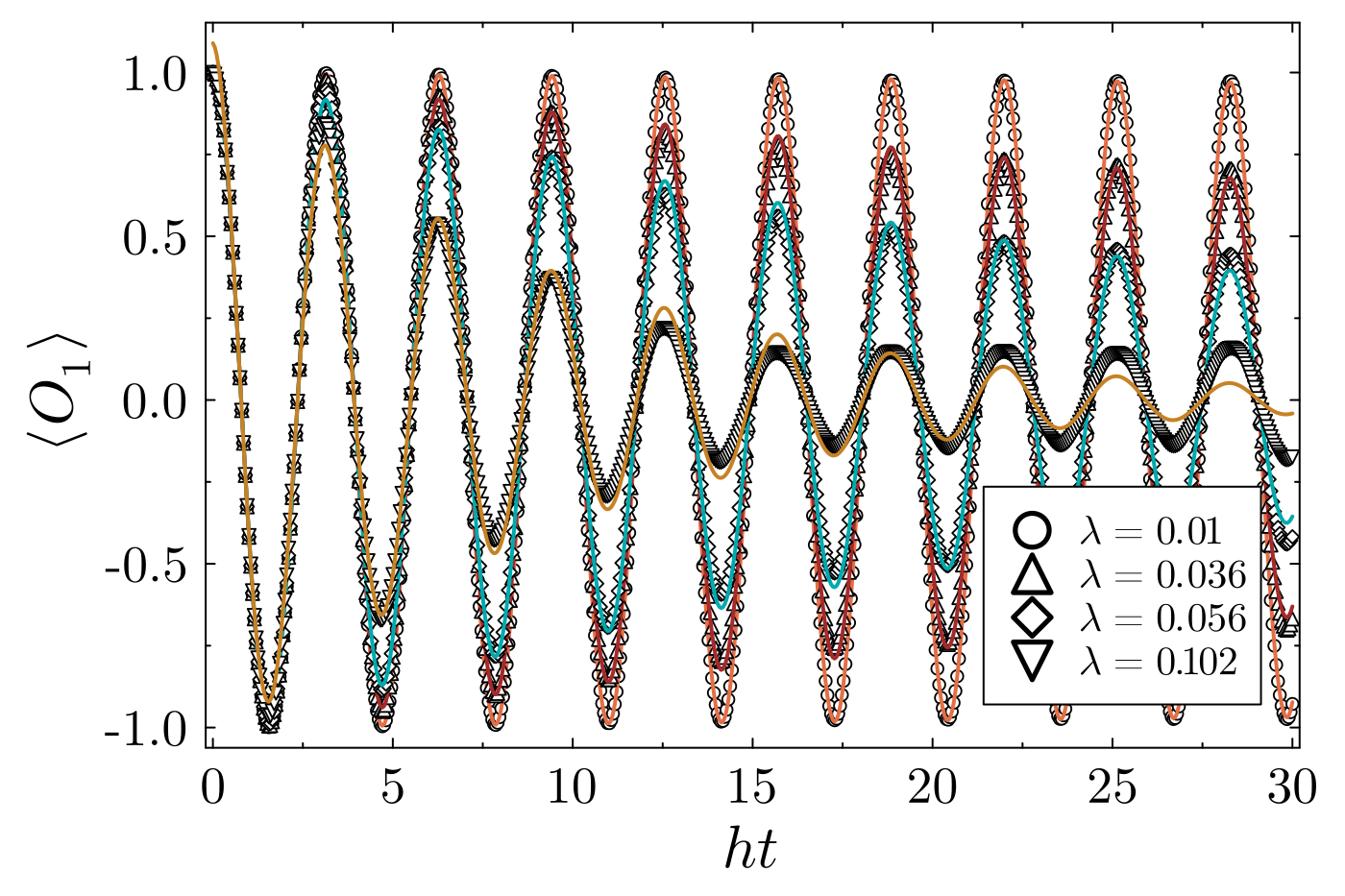}
        \put(17,59){\large (a)}
        \end{overpic}
        \phantomcaption\label{fig:ptb_director}
    \end{subfigure}%
    \\[-4mm]
    \begin{subfigure}{\columnwidth}
        \centering
        \begin{overpic}[width=\columnwidth]     {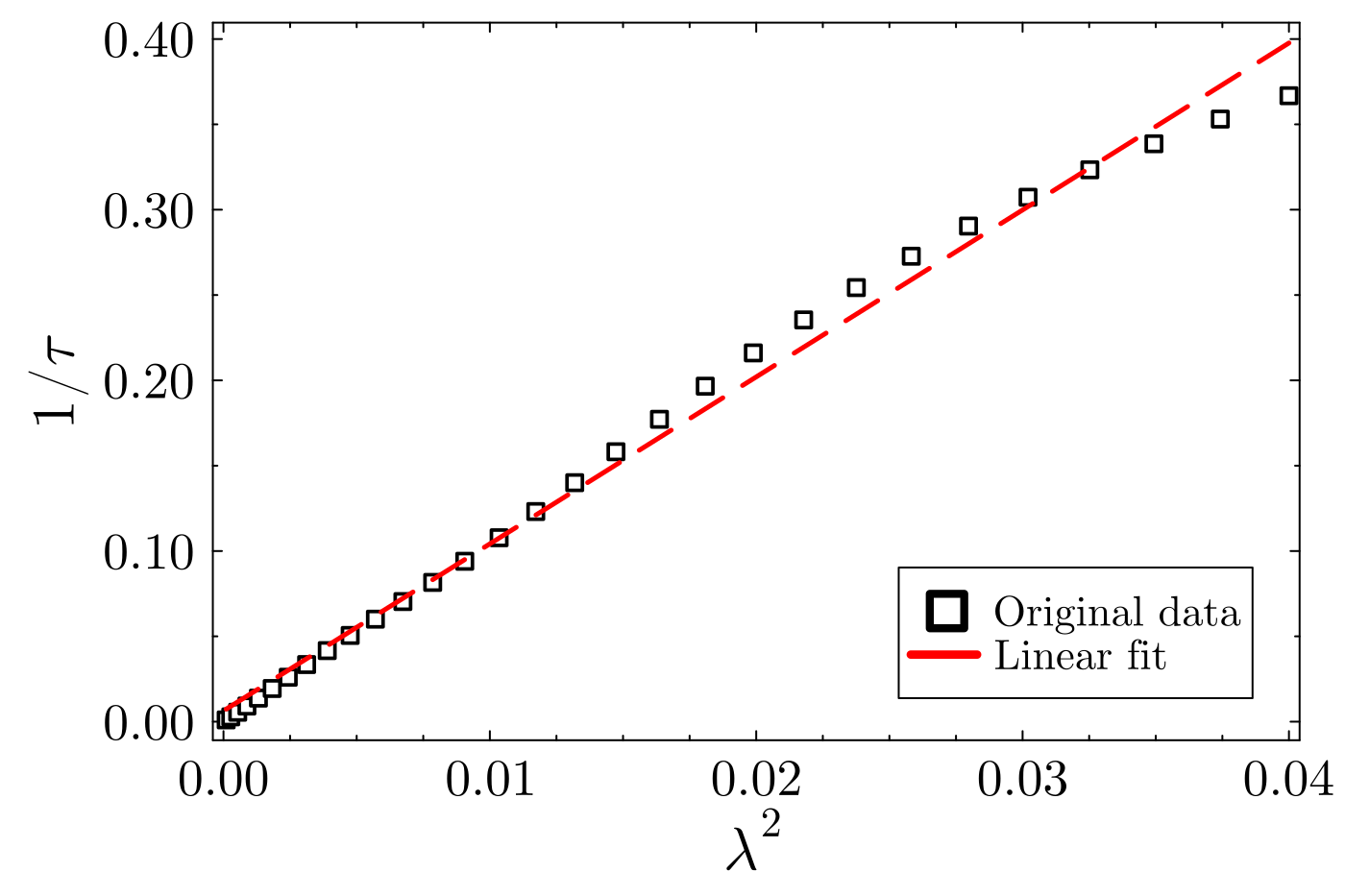}
        \put(17,59){\large (b)}
        \end{overpic}
        \phantomcaption\label{fig:ptb_director_fit}
    \end{subfigure}
    \caption{(a) Time evolution of $\braket{O_1}$ following a quench from the initial state $\ket{\psi_0}$ Eq.~\eqref{neel_state} for different perturbation strengths $\lambda$, for a chain of length $L=8$ with rest of the parameters same as in Fig.~\ref{fig:unp_entang}. The decay lifetime $\tau$ is obtained by fitting the oscillation envelope to the function $\braket{O_1(t)}=Ae^{-t/\tau}\cos(\omega t)$. The markers denote the original numerical results and the solid lines represent the exponential fits. (b) Dependence of the inverse lifetime ($\tau^{-1}$) on squared perturbation strength ($\lambda^2$).}
    \label{fig:director_decay}
\end{figure}
\noindent gradually lose their athermal properties in presence of next nearest-neighbor exchange. The revivals persist over long-time evolution for small values of $\lambda$, but they rapidly disappear as the perturbation strength increases. The first peak of fidelity $\mathcal{F}(t)$ of the initial N\'eel state decreases approximately as a quadratic function of $\lambda^2$, see Fig.~\ref{fig:fidelity_decay_fit}. 

The characteristic long-range order (ODLRO) of the scarred states, which distinguish these special eigenstates from ETH-following states, is also suppressed in presence of the perturbation. As presented in Fig.~\ref{fig:ptb_odlro}, the connected correlations decrease upon adding the perturbation and continue to diminish with increasing strength of the next nearest-neighbor exchange interaction. The ODLRO shown in the plot is calculated for the descendant state of an unperturbed scarred state, identified as the eigenstate in the perturbed manifold that has the maximum overlap with the corresponding scarred state. It should be noted that as the perturbation strength $\lambda$ increases, the scarred states hybridize more strongly with nearby eigenstates, which makes it increasingly less practical to pick out a single eigenstate as the descendant state. Nonetheless, the decay of the long-range order points to the eventual thermalization of the scarred states as an ETH-obeying eigenstate in the middle of the spectrum, which is essentially viewed as a random vector, is not expected to exhibit long-range correlations.

We now investigate how the perturbation modifies the perfect oscillations of the local observable defined in Eq.~\eqref{nematic_director_param} for the nematic N\'eel state. As shown in Fig.~\ref{fig:ptb_director}, the oscillations persist at small perturbation strengths $\lambda$, but they gradually damp out as $\lambda$ increases. The envelope of the decaying oscillation of the expectation value of the spin-nematic order parameter, $\braket{O_1(t)}$, can be captured by an exponentially decaying function, $y(t)=Ae^{-t/\tau}\cos(\omega t)$, where $A$, $\tau$, $\omega$ are the fitting parameters. In the weak perturbation regime, this exponential form accurately describes the damping of the oscillation. However, as the next nearest-neighbor interaction becomes stronger, the function fails to describe the oscillation at late times. The parameter $\tau$ characterizes the lifetime of the oscillation, after which the thermalization effects become dominant. As Fig.~\ref{fig:ptb_director_fit} reveals, the inverse lifetime varies approximately linearly with the perturbation strength ($\tau^{-1}\sim\lambda^2$) for small values of $\lambda$, which is consistent with Fermi's golden rule \cite{fermi_golden_iadecola,fermi_golden_rigol_1,fermi_golden_rigol_2}. 

\section{Conclusions}
In summary, the central objective of this work is to examine the stability of QMBSs against perturbation in the spin-1 XY chain. The tower of scarred states constructed from bimagnon excitations at momentum $\pi$ forms a set of exact eigenstates of the unperturbed XY model. The nonthermalizing nature of these special states is manifested through long-range correlations and late-time coherent oscillations following a quench from an initial state residing within the scarred subspace.

When a next nearest-neighbor exchange interaction is introduced, the QMBSs begin to hybridize with ETH-obeying eigenstates in the spectrum. For small chains, the deformed scar states can still be perturbatively associated with the exact scarred states for small perturbation strengths. However, from the scaling of the off-diagonal perturbation matrix elements, it can be argued that the scars will gradually thermalize due to strong hybridization with thermal eigenstates as the length of the chain increases. Thus, strong mixing makes these scarred states inaccessible using perturbation theory in the thermodynamic limit. The disappearing features of the QMBSs under perturbation are reflected in the diminishing off-diagonal long range order and the damping of oscillations in certain local observables. Furthermore, in the weak perturbation regime, the thermalization time is found to scale as $1/\lambda^2$ with the perturbation strength $\lambda$.

\section*{Acknowledgments}
The author thanks Sinchan Ghosh, Bhaskar Mukherjee, Anubhab Rudra, and Krishnendu Sengupta for discussions and helpful suggestions. The author acknowledges supports from Indian Association for the Cultivation of Science in terms of resources and computing facilities.

\appendix
 \titleformat{\section}
  {\bfseries\centering\uppercase}{}{0pt}{\MakeUppercase{APPENDIX \thesection:\hspace{0.4em}}}[] 
   
\section{Spectrum Generating Algebra}
\label{appendix:SGA}
The nonthermal special eigenstates of the spin-1 \textit{XY} model presented in Eq.~\ref{first_NN_scar} are part of a spectrum generating algebra (SGA) which was originally introduced in high energy physics \cite{sga_high_energy} and later applied to the Hubbard model \cite{sga_chen,sga_zhang,sga_moud}. After some calculations, it can be shown that for the \textit{XY} Hamiltonian $H_0$ and the bimagnon creation operator $J^+$, the following relation holds,
\begin{align}
\left(\left[H_0,J^+\right]\right)W=\left(2hJ^+\right)W
\label{app_eq:SGA_eq}
\end{align}
\noindent where $W$ is the scarred subspace, and $h$ is the magnetic field. It is important to note that the above relation is restricted only to the scarred subspace and is not valid for the entire Hilbert space. These properties, together with the fact that $J^+$ is not related to any symmetry of the Hamiltonian, are instrumental for the athermal properties of the scarred eigenstates.

Given that Eq.~\eqref{app_eq:SGA_eq} holds and $\ket{\Omega}$ is the fully polarized vacuum state belonging to $W$, which is an eigenstate of $H_0$ with energy $E_0$, and states created by the operator $J^+$ stay within the scarred subspace (i.e. $J^+W\subset W$), the family of states,
\begin{align}
    \ket{\mathcal{S}{}_n}=\left(J^+\right)^n\ket{\Omega}
\end{align}
\noindent are also eigenstates of $H_0$ with energies $E_0+2hn$, provided $\left(J^+\right)^n\ket{\Omega}\neq0$ \cite{sga_lin}. For the spin-1 \textit{XY} model described in Eq.~\eqref{H_0}, $E_0$ is $(D-L)h$. Thus SGA, which is responsible for the decoupling of the non-thermalizing subspace from the thermal subspace \cite{weak_ergod}, also explains the equal spacing of the scarred states in the spectrum.

\section{Additional results on hybridization and matrix-element scaling}
\label{appendix:add_overlap_scaling}
\begin{figure}[H]
    \centering 
    \begin{subfigure}{0.49\columnwidth} 
        \centering 
        \begin{overpic}[width=\linewidth] {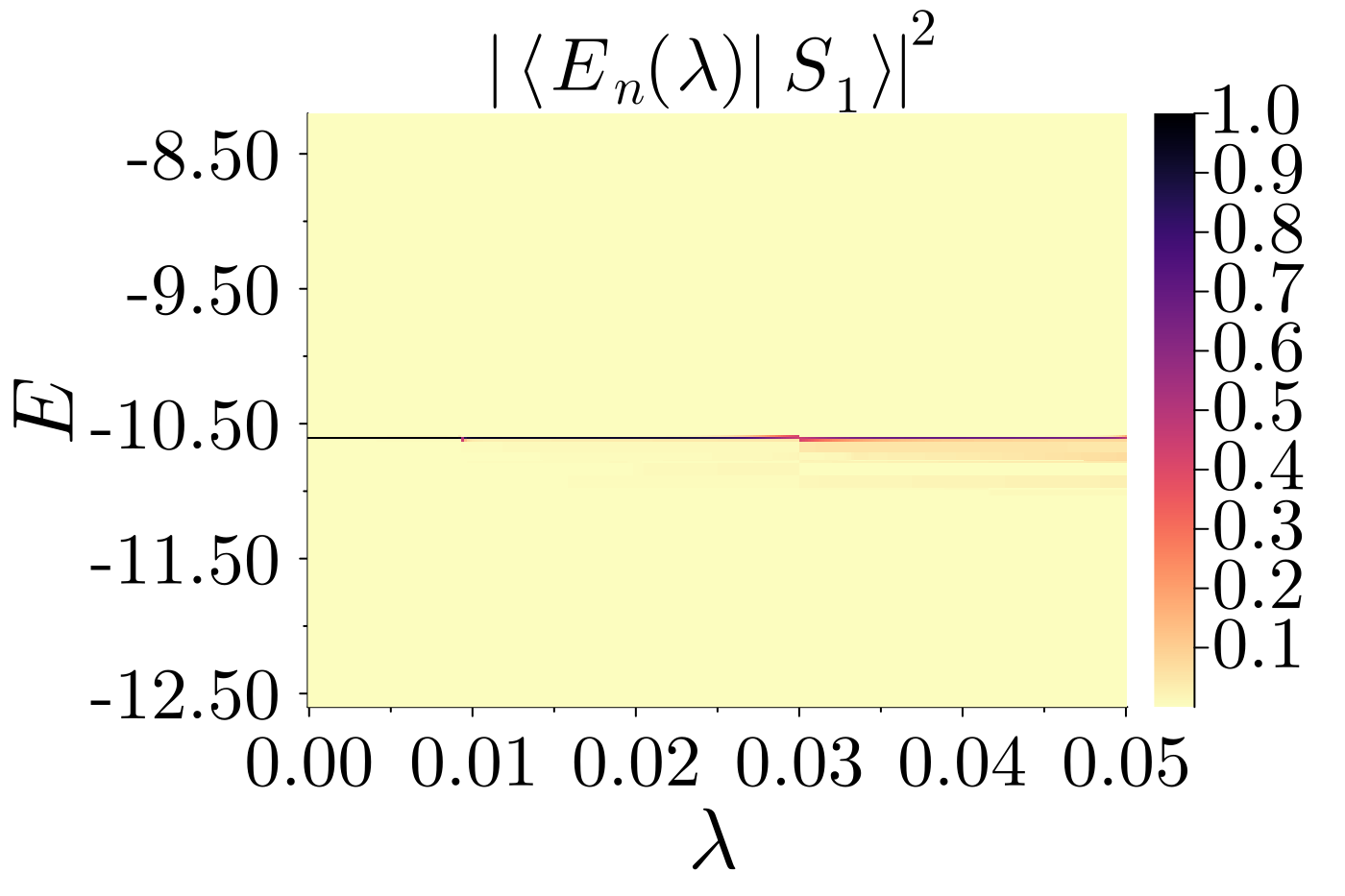} 
        \put(25,50){(a)} 
        \end{overpic} 
        \phantomcaption
        \label{fig:add_S1_overlap} 
    \end{subfigure} 
    \hspace{-0.9mm}
    \begin{subfigure}{0.49\columnwidth} 
        \centering 
        \begin{overpic}[width=\linewidth] {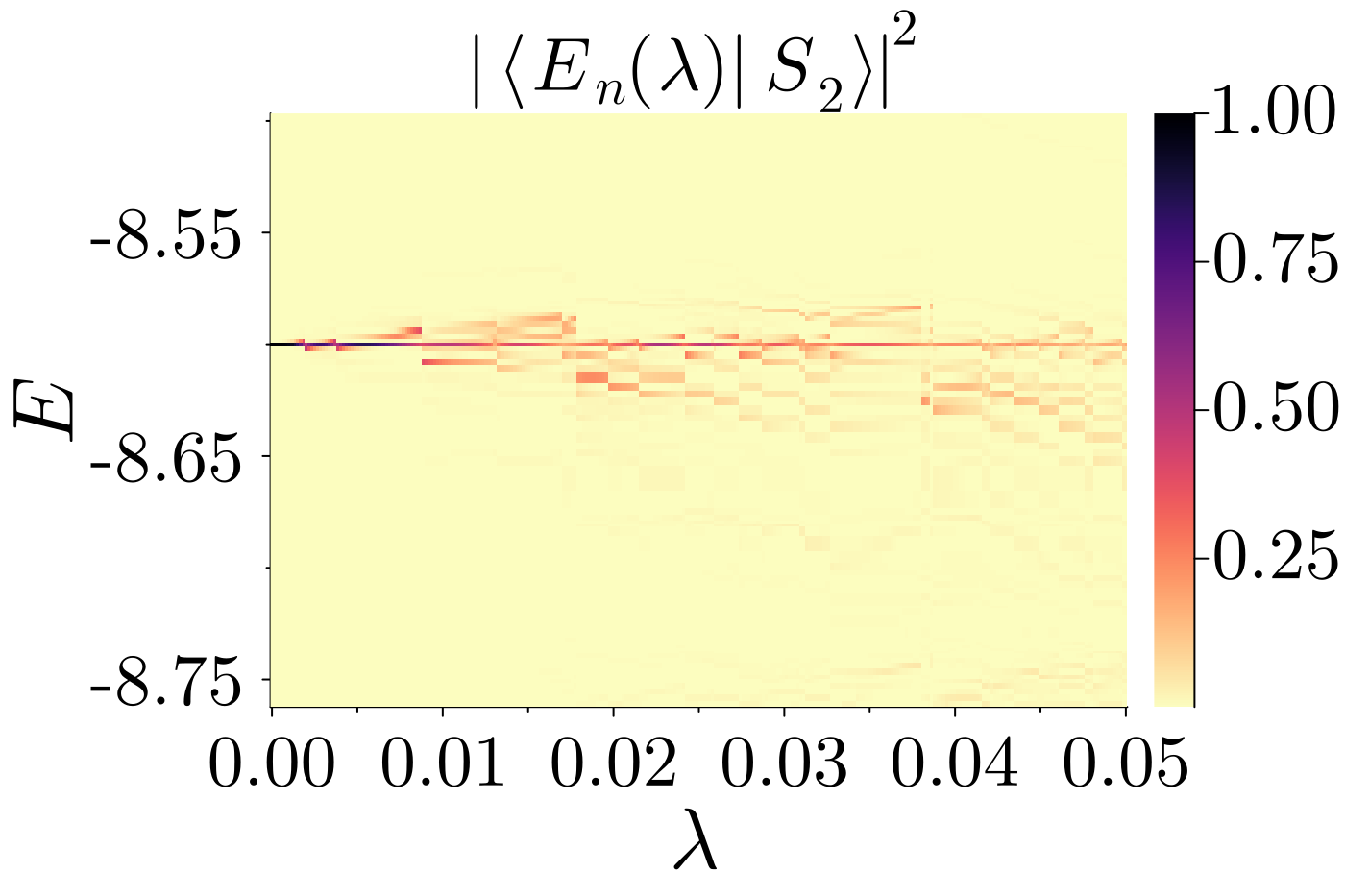} 
        \put(25,50){(b)} 
        \end{overpic} 
        \phantomcaption
        \label{fig:add_S2_overlap} 
    \end{subfigure} 
    \\[-4mm]
    \begin{subfigure}{0.49\columnwidth} 
        \centering 
        \begin{overpic}[width=\linewidth] {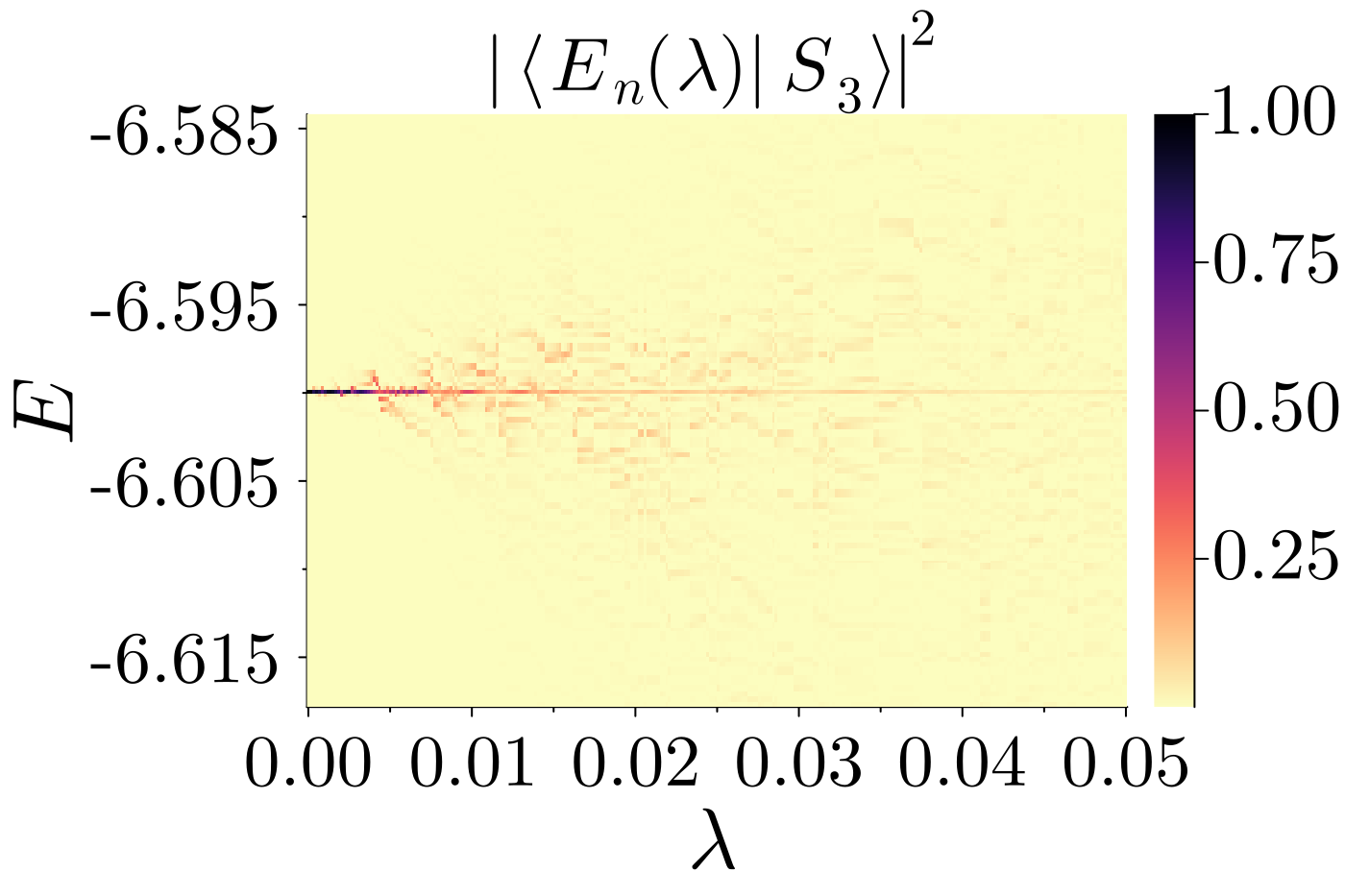} 
        \put(25,50){(c)} 
        \end{overpic} 
        \phantomcaption
        \label{fig:add_S3_overlap} 
    \end{subfigure} 
    \hspace{-0.9mm}
    \begin{subfigure}{0.49\columnwidth} 
        \centering 
        \begin{overpic}[width=\linewidth] {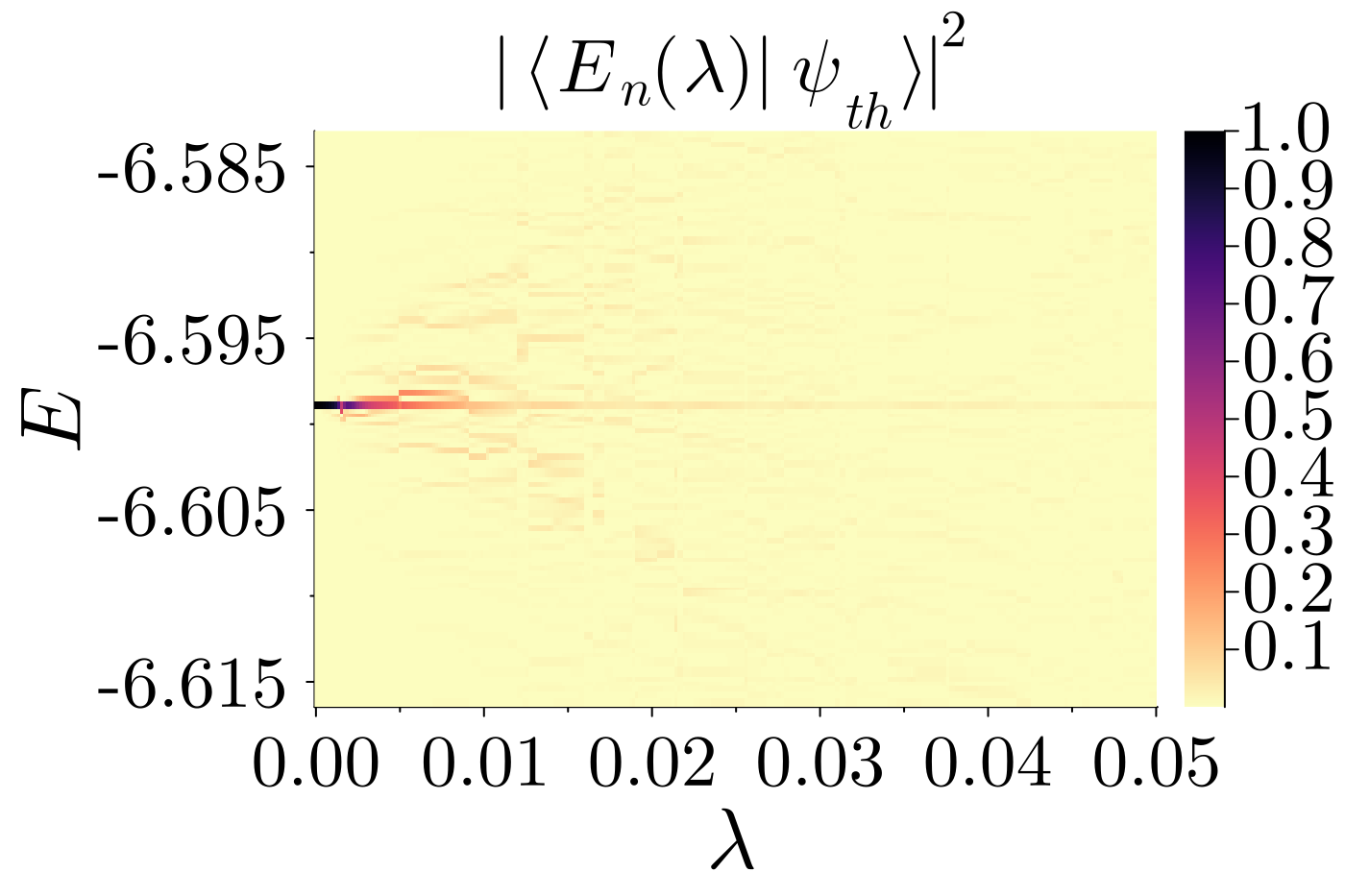} 
        \put(25,50){(d)} 
        \end{overpic} 
        \phantomcaption
        \label{fig:add_th_overlap} 
    \end{subfigure} 
\caption{\justifying Heatmap plot of squared overlaps of the exact eigenstates $|E_n(\lambda)\rangle$ of the perturbed Hamiltonian $H$ Eq.~\eqref{H} for $L=14$, while the remaining parameters are same as in Fig.~\ref{fig:unp_entang}, with: the scarred states (a), (b), (c) the $|\mathcal{S}{}_1\rangle$, $|\mathcal{S}{}_2\rangle$, $|\mathcal{S}{}_3\rangle$ respectively, and (d) the thermal state $|\psi_{th}\rangle$. The thermal state is chosen as the eigenstate of the unperturbed \textit{XY} Hamiltonian with eigenindex four more than the index of $|\mathcal{S}{}_3\rangle$ in the same $S^z_{\mathrm{tot}}$ sector.}
\label{fig:add_scar_th_overlap} 
\end{figure}

As shown in Fig.~\ref{fig:add_scar_th_overlap}, the $\ket{\mathcal{S}{}_1}$ state displays relatively higher robustness against perturbation for an $L=14$ chain compared to other scarred states considered in this study. The weights of the $\ket{\mathcal{S}{}_2}$ state become significantly distributed over many eigenstates even at small perturbation strengths. On the other hand, the scarred state $\ket{\mathcal{S}{}_3}$ practically shows almost no distinction from the thermal state in its hybridization behavior. The overlaps of both states with other eigenstates spread so broadly across the perturbed manifold that beyond $\lambda\approx0.01$ it becomes im-
\noindent possible to identify the original state. For comparison, in a smaller chain of length $L=6$, the same scarred states exhibit noticeably weaker hybridization over the same range of perturbation strengths, see Fig.~\ref{fig:scar_th_overlap}. The enhanced mixing with eigenstates of the perturbed Hamiltonian as the system size increases supports the conclusion that the nonthermal features of the scarred states eventually disappear.

Fig.~\ref{fig:S2_pert_scaling_ediff} illustrates the averaged matrix element coupling $\ket{\mathcal{S}{}_2}$ to eigenstates with absolute energy differences 
{\setlength{\belowcaptionskip}{0pt}
\begin{figure}[H]
    \centering 
    \begin{subfigure}{0.487\columnwidth} 
        \centering 
        \begin{overpic}[width=\linewidth] {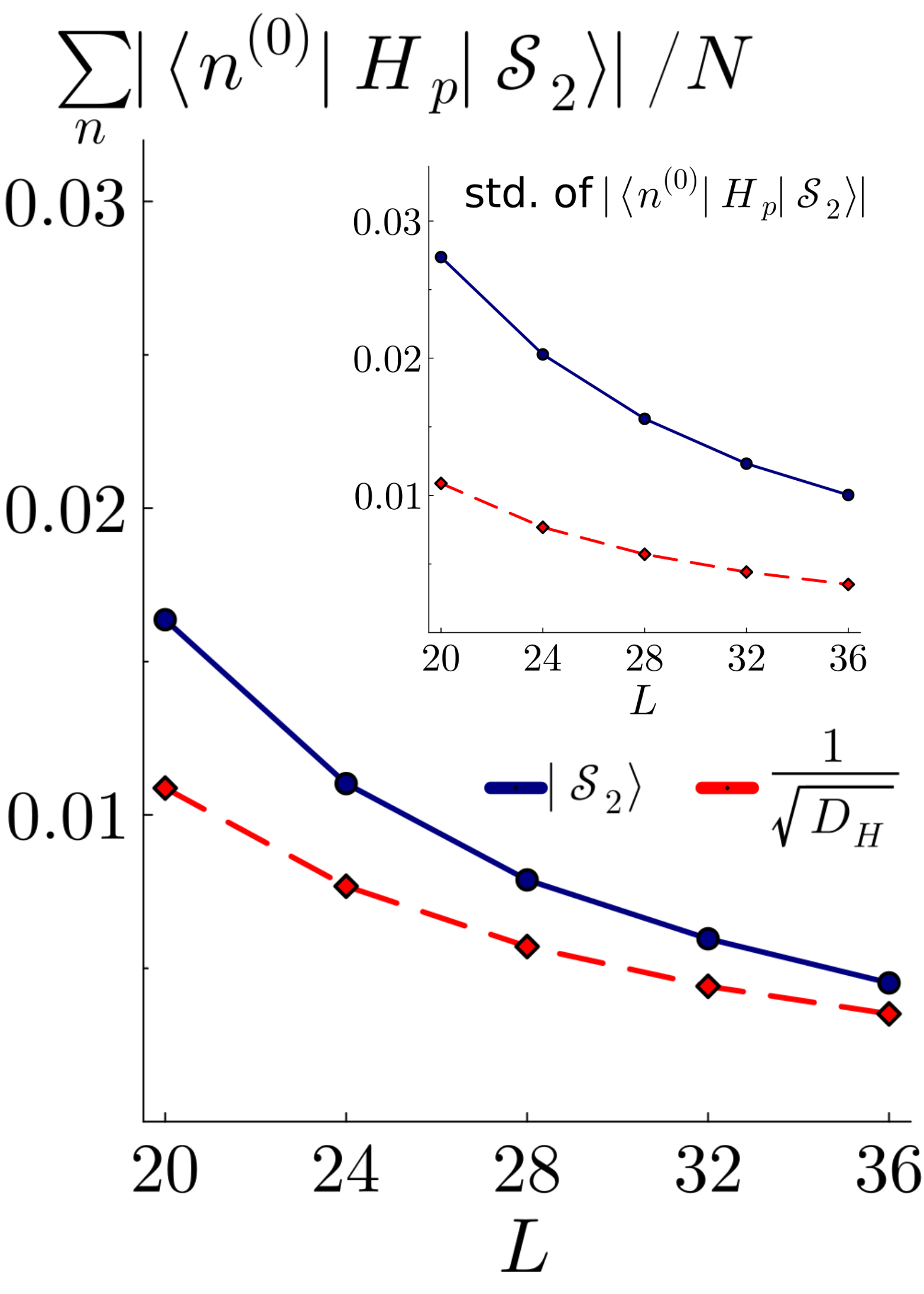} 
        \put(15,18){(a)} 
        \end{overpic} 
        \phantomcaption
        \label{fig:S2_pert_scaling_ediff} 
    \end{subfigure} 
    \hspace{0.00001mm}
    \begin{subfigure}{0.487\columnwidth} 
        \centering 
        \begin{overpic}[width=\linewidth] {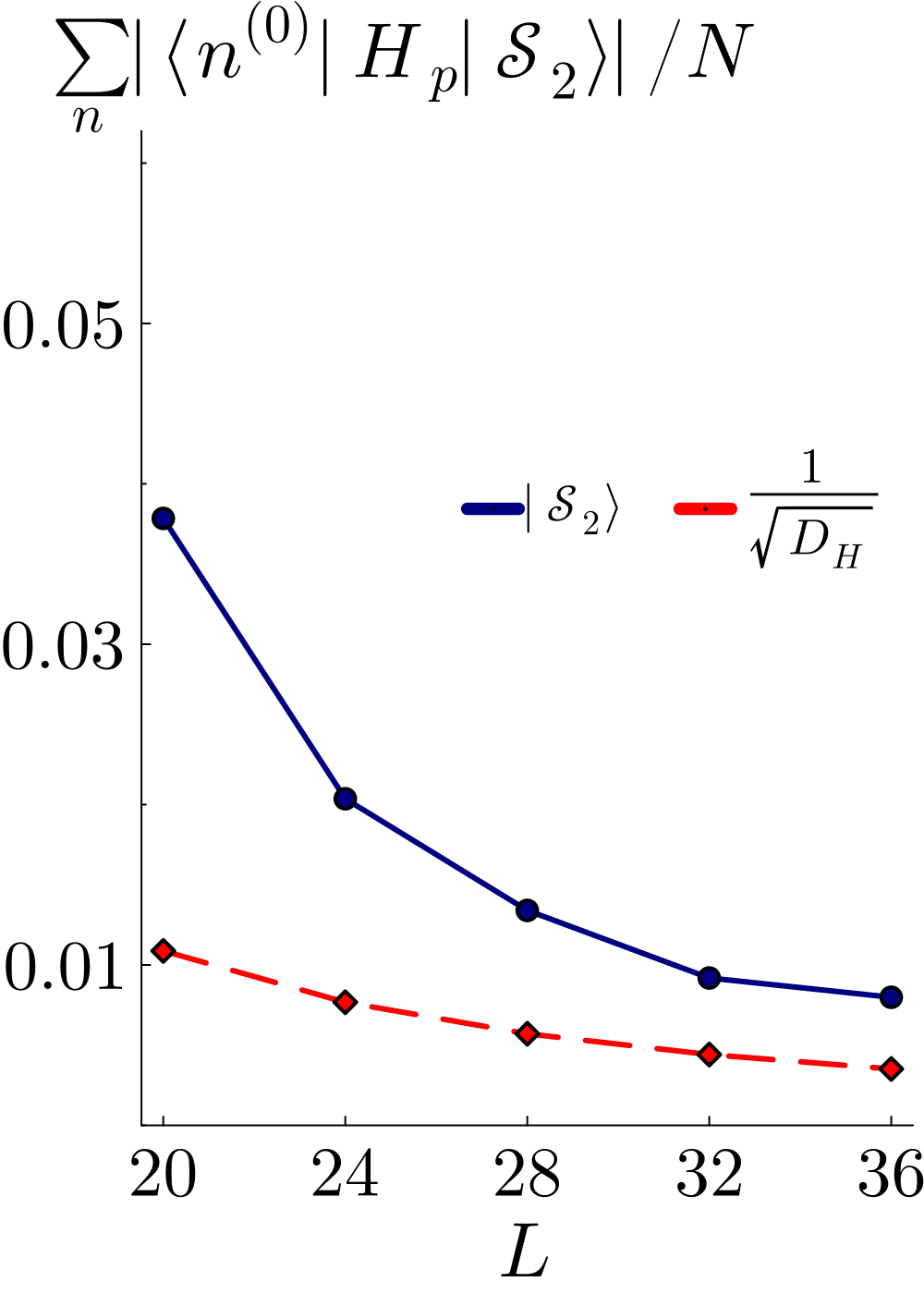} 
        \put(15,18){(b)} 
        \end{overpic} 
        \phantomcaption
        \label{fig:S2_pert_scaling_smallest} 
    \end{subfigure}
    \caption{Finite-size scaling of the matrix elements for the scarred state $\ket{\mathcal{S}{}_2}$: (a) average of $|\langle n^{(0)}|H_p|\mathcal{S}{}_2\rangle|$ over the eigenstates $\ket{n^{(0)}}$ such that $|E_n^{(0)}-E_{\mathcal{S}{}_2}^{(0)}|<3$. Inset shows scaling of standard deviation of $|\langle n^{(0)}|H_p|\mathcal{S}{}_2\rangle|$ calculated within the same energy window. (b) $|\langle n^{(0)}|H_p|\mathcal{S}{}_2\rangle|$ averaged over $800$ eigenstates with the smallest $|E_n^{(0)}-E_{\mathcal{S}{}_2}^{(0)}|$.}
    \label{fig:S2_pert_element_scaling}
\end{figure}}
\begin{figure*}[t!]
    \centering 
    \begin{subfigure}{0.325\textwidth} 
        \centering 
        \begin{overpic}[width=\linewidth] {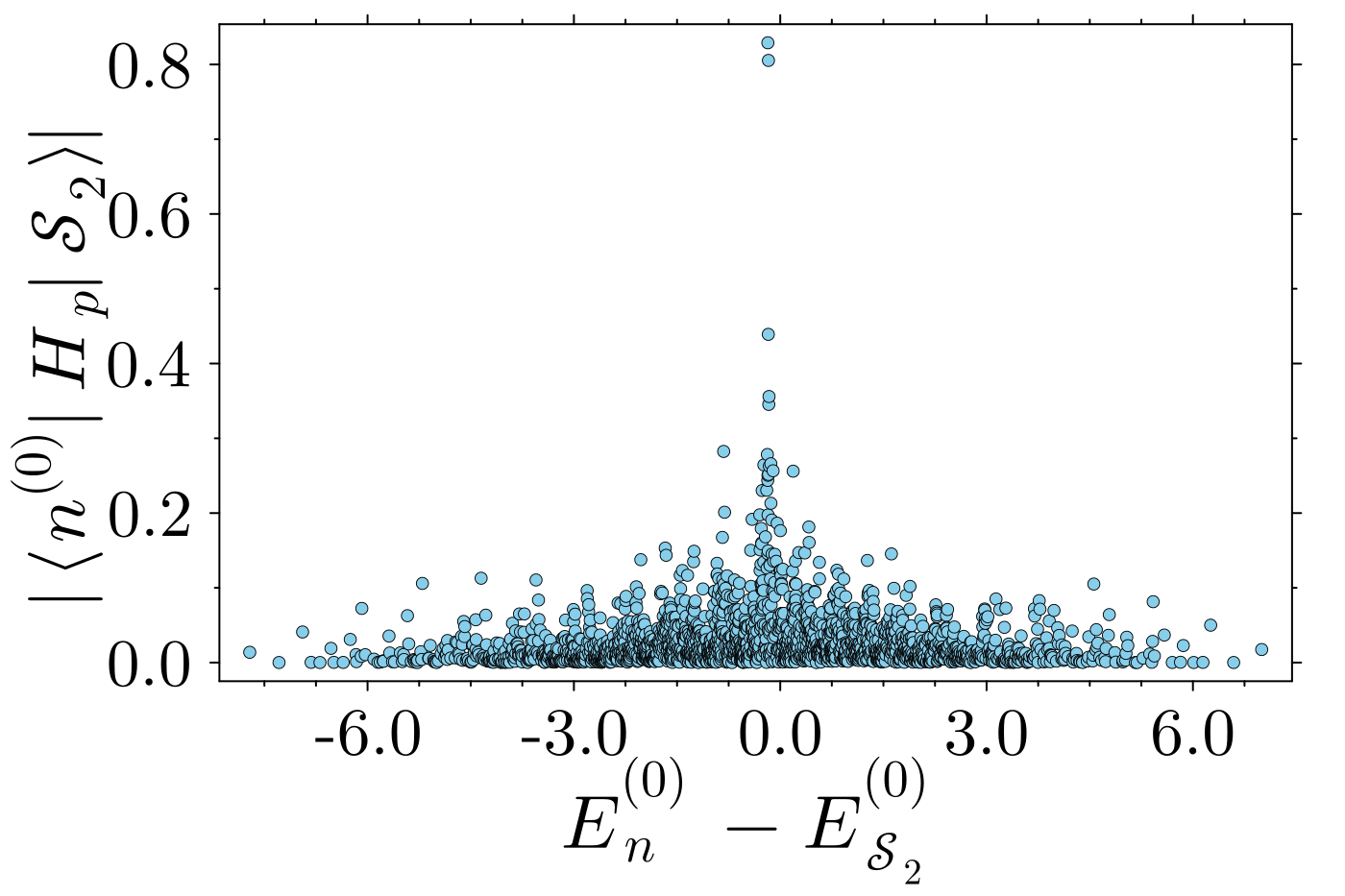} 
        \put(19,54){(a)} 
        \end{overpic} 
        \phantomcaption
        \label{fig:S2_pert_dist} 
    \end{subfigure} 
    \hspace{0.005mm}
    \begin{subfigure}{0.325\textwidth} 
        \centering 
        \begin{overpic}[width=\linewidth] {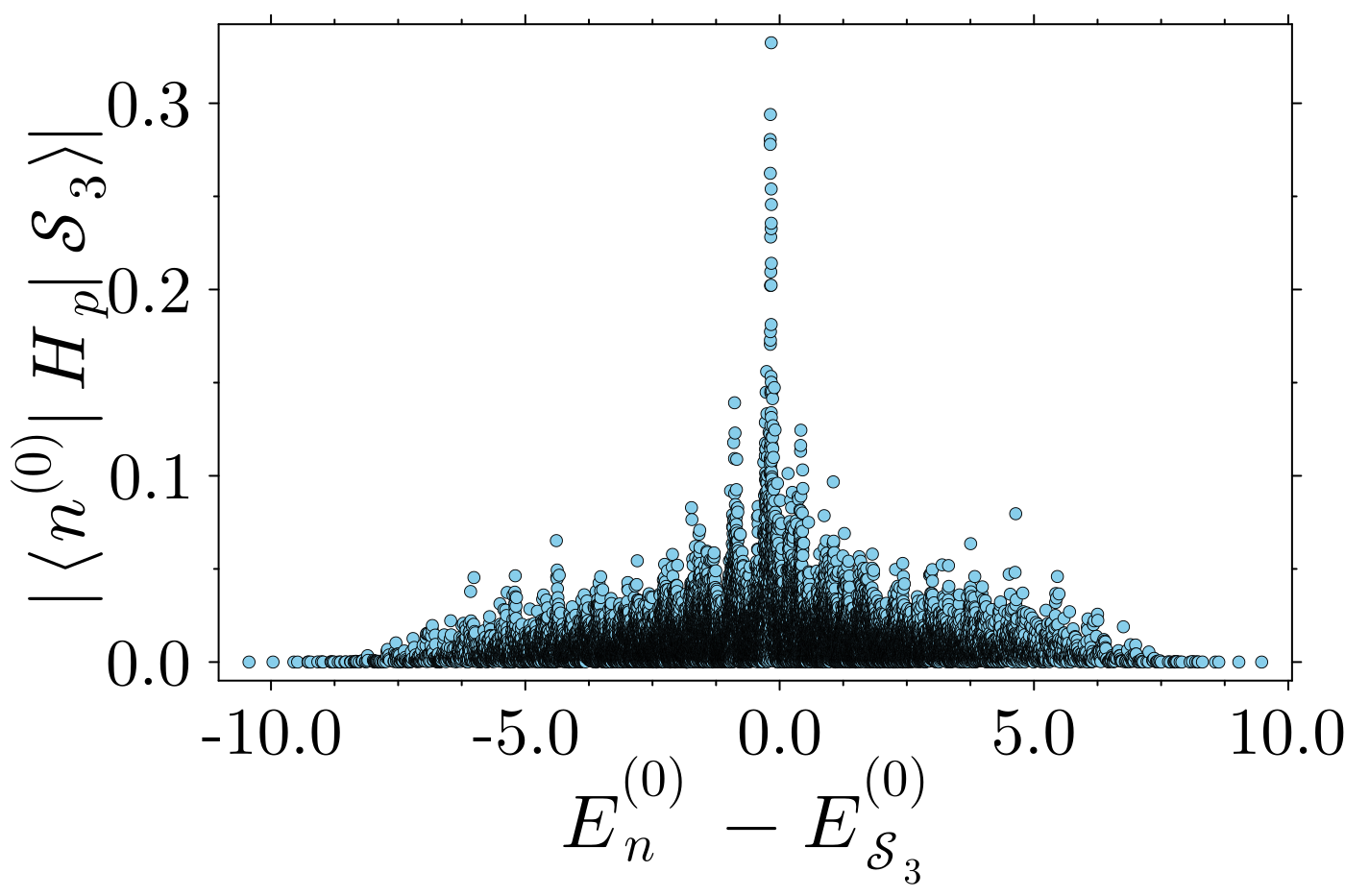} 
        \put(19,54){(b)} 
        \end{overpic} 
        \phantomcaption
        \label{fig:S3_pert_dist} 
    \end{subfigure} 
    \hspace{0.005mm}
    \begin{subfigure}{0.325\textwidth} 
        \centering 
        \begin{overpic}[width=\linewidth] {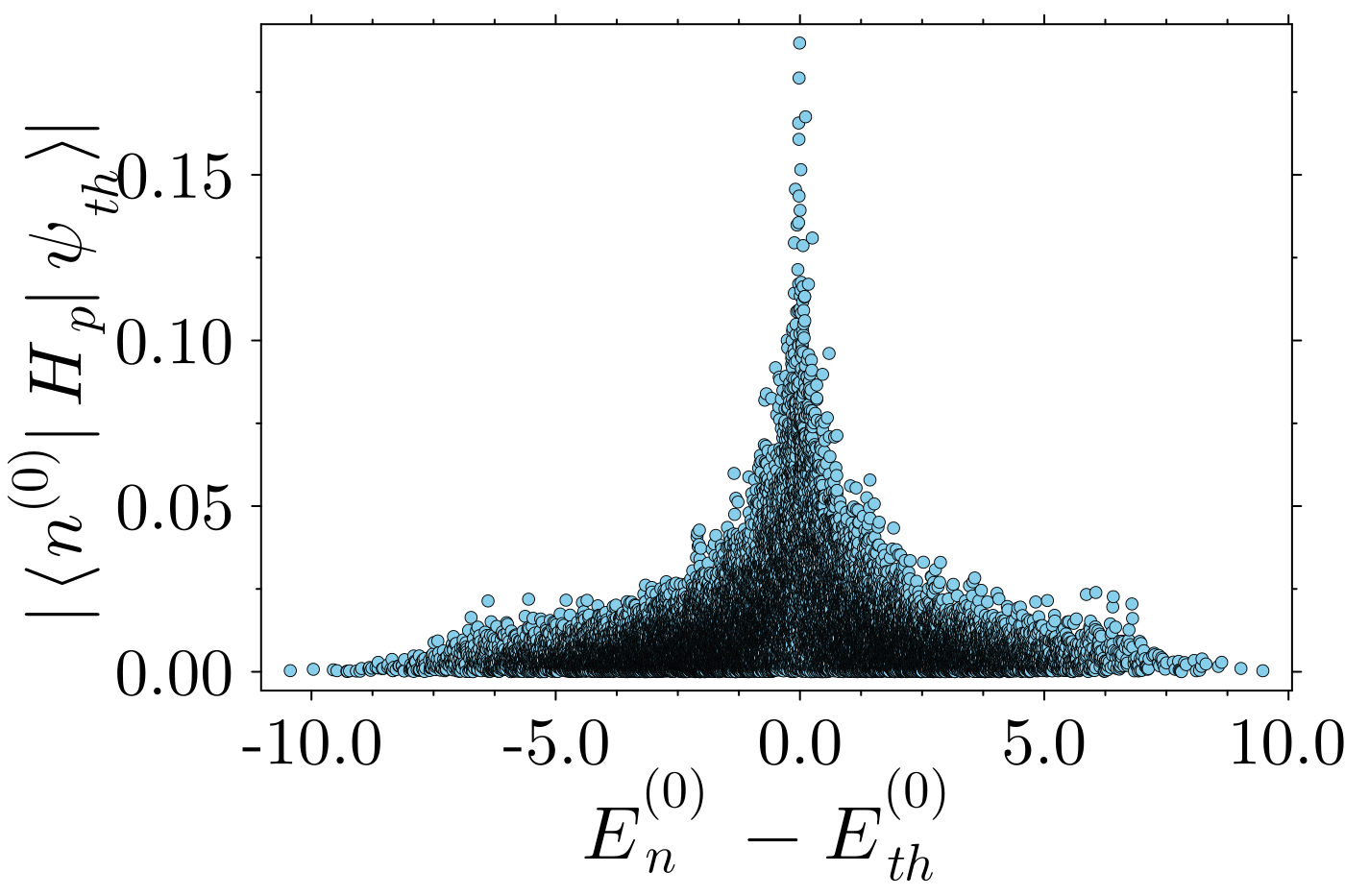} 
        \put(22,54){(c)} 
        \end{overpic} 
        \phantomcaption
        \label{fig:thermal_pert_dist} 
    \end{subfigure} 
\caption{\justifying Distribution of the perturbation matrix elements $|\langle n^{(0)}|H_p|\psi\rangle|$, where $\ket{\psi}$ is (a) the scarred state $\ket{\mathcal{S}{}_2}$, (b) the scarred state $\ket{\mathcal{S}{}_3}$, and (c) the thermal state $\ket{\psi_{th}}$, for an $L=14$ chain with the other parameters same as in Fig.~\ref{fig:unp_entang}. The thermal state is chosen as the eigenstate of the \textit{XY} Hamiltonian $H_0$ whose eigenindex is four greater than that of $\ket{\mathcal{S}{}_3}$ in the same $S^z_{\mathrm{tot}}$ sector.}
\label{fig:pert_matrix_element_dist} 
\end{figure*}
\noindent smaller than $3$, and Fig.~\ref{fig:S2_pert_scaling_smallest} shows the corresponding average when taken over $800$ states having the smallest energy differences. It is observed that the averaged matrix element for the two-bimagnon state also approximately follows the $D_H^{-1/2}$ scaling for both types of averaging, similar to $\ket{\mathcal{S}{}_3}$ state shown in Fig.~\ref{fig:S3_pert_element_scaling}. For the $\ket{\mathcal{S}{}_2}$ state in Fig.~\ref{fig:S2_pert_element_scaling}, a range of larger system sizes has been chosen to ensure that a sufficient number of eigenstates are available within the $S^z_{\mathrm{tot}}$ sector, so that the Hilbert space dimension is reasonably larger than the fraction of states satisfying the averaging conditions.  

\section{Distribution of Perturbation matrix elements}
\label{appendix:dist_pert_matrix_elem}
Fig.~\ref{fig:pert_matrix_element_dist} presents the distribution of the perturbation matrix elements $H_p$, which quantify how $H_p$ couples different scarred states and the thermal state to eigenstates at different energies in the spectrum. For the thermal state $\ket{\psi_{th}}$, the distribution is relatively uniform and it shows a prominent peak for eigenstates whose energies are very close to that of $\ket{\psi_{th}}$. A similar peak is observed for the scarred states as well at a small but finite energy difference, along with additional smaller peaks appearing at higher energy differences. Overall, the distributions for the scarred states and the thermal state share the same qualitative structure, although the scarred states exhibit a narrower central peak. For both types of states, the matrix elements have larger amplitudes at small energy differences, and their magnitudes decrease nearly uniformly for eigenstates at larger energy separations. This behavior indicates strong hybridization with nearby eigenstates in the spectrum, which is also evident in Fig.~\ref{fig:add_scar_th_overlap}, where the overlaps highly spread across many eigenstates. Nonetheless, the additional features of the scar states suggest a more selective coupling under $H_p$.

\bibliography{references}
\end{document}